\documentclass[sigconf]{acmart}

\AtBeginDocument{%
  \providecommand\BibTeX{{%
    \normalfont B\kern-0.5em{\scshape i\kern-0.25em b}\kern-0.8em\TeX}}}

\setcopyright{acmcopyright}
\copyrightyear{2022}
\acmYear{2022}
\acmDOI{10.1145/1122445.1122456}

\acmConference[CHI '22]{CHI '22: ACM CHI Conference on Human Factors in Computing Systems}{April 30--May 06, 2022}{New Orleans, LA}
\acmBooktitle{CHI '22: CHI Conference on Human Factors in Computing Systems,
  April 30--May 06, 2022, New Orleans, LA}
\acmPrice{15.00}
\acmISBN{978-1-4503-XXXX-X/18/06}

\newcommand\subjectcount{246}

\newcommand{\todo}[1]{{\color{black} #1}}

\usepackage{multirow}
\usepackage{subcaption}
\usepackage{caption}
\begin{document}

\title[TalkTive: A Conversational Agent Using Backchannels to Engage Older Adults]{TalkTive: A Conversational Agent Using Backchannels to Engage Older Adults in Neurocognitive Disorders Screening}

\author{Zijian Ding}
\affiliation{%
  \institution{College of Information Studies, University of Maryland College Park}
  \country{USA}}
 
\author{Jiawen Kang}
\affiliation{%
  \institution{Department of SEEM, The Chinese University of Hong Kong}
  \country{HKSAR}}
  
\author{Tinky Oi Ting HO}
\affiliation{%
  \institution{Department of Psychology, The Chinese University of Hong Kong}
  \country{HKSAR}}
  
\author{Ka Ho Wong}
\affiliation{%
  \institution{Department of SEEM, The Chinese University of Hong Kong}
  \country{HKSAR}}
  
\author{Helene H. Fung}
\affiliation{%
  \institution{Department of Psychology, The Chinese University of Hong Kong}
  \country{HKSAR}}

\author{Helen Meng}
\affiliation{%
  \institution{Department of SEEM, The Chinese University of Hong Kong}
  \country{HKSAR}}
  
\author{Xiaojuan Ma}
\affiliation{%
  \institution{Department of CSE,  Hong Kong University of Science and Technology}
  \country{HKSAR}}

\renewcommand{\shortauthors}{Ding, et al.}

\begin{abstract}
 Conversational agents (CAs) have the great potential in mitigating the clinicians' burden in screening for neurocognitive disorders among older adults. It is important, therefore, to develop CAs that can be engaging, to elicit conversational speech input from older adult participants for supporting assessment of cognitive abilities. As an initial step, this paper presents research in developing the backchanneling ability in CAs in the form of a verbal response to engage the speaker. We analyzed 246 conversations of cognitive assessments between older adults and human assessors, and derived the categories of reactive backchannels (e.g. ``hmm”) and proactive backchannels (e.g. ``please keep going”). This is used in the development of \textit{TalkTive}, a CA which can predict both timing and form of backchanneling during cognitive assessments. The study then invited 36 older adult participants to evaluate the backchanneling feature. Results show that proactive backchanneling is more appreciated by participants than reactive backchanneling.
\end{abstract}





\maketitle

\def \RQO {\textbf{RQ1}: When do reactive and proactive backchannels happen in a task-driven conversation with older adults in Cantonese?}

\def \RQT {\textbf{RQ2}: How do reactive and proactive backchannels provided by a task-oriented CA affect its conversation with older adults in Cantonese?}

\section{Introduction}
\label{sec:intro}
The rapidly ageing global population is imposing challenges to healthcare systems across nations \cite{department2019world}. Neurocognitive disorders (NCD), such as dementia, are particularly common in older adults \cite{alzheimer20192019}. The global cost of NCD exceeded the threshold of US\$1 trillion in 2018 \cite{patterson2018world}. In Hong Kong, the estimated cost of institutional and informal care for older adults with NCD in 2036 is HK\$31.2 billion (US\$4 billion) \cite{yu2010dementia}. \todo{Besides leading to an enormous financial burden on the society, NCD has a negative impact on the quality of life of older adults, their families, and caretakers, as well as on the workload of clinicians  \cite{patterson2018world}}.
% Kenji Toba, ``A third of babies born now in Japan will live to 100 years. The risk of dementia in a
% centenarian in Japan is 99\%. Everyone has to understand
% `it's my story'. Not your story. `Cognitive decline is my
% story.'" 
To look on the bright side, the negative symptoms associated with NCD have been shown to be controllable if patients can get access to early diagnoses and timely preventive interventions \cite{akpan2019neurocognitive, ferri2005global}. This presents a necessity for a scalable approach in screening cognitive impairments among older adults. Current NCD screening and diagnosis tests, such as the Montreal Cognitive Assessment (MoCA) \cite{nasreddine2005montreal}, are mainly conducted as in-person tests by clinical professionals  \cite{tang2020cccdtd5, ye2021development}. In-person assessments face limitations due to various factors, such as limited accessibility to the tests due to the lower mobility of some older adults, and shortages and inter-rater variabilities of clinicians \cite{lucza2015screening}.

% pandemic is not permanent - social distancing restrictions amidst the global pandemic \cite{fatmi2020covid},
 
To overcome these limitations, researchers are working on alternative solutions for NCD screening. Since NCDs are often manifested as communicative impairments, machine learning (ML) algorithms are offering a new type of support for NCD screening \cite{pan2019automatic, mirheidari2019dementia, li2021comparative, fraser2016linguistic, weiner2016speech, al2016simple, mirheidari2018detecting, konig2018fully, pompili2020pragmatic, chakraborty2020identification, ye2021development}. Therefore, a potential solution is to integrate speech analytics into a conversational agent (CA) to support highly accessible voice-based web applications that can interact with older adults through spoken language for screening NCD. Older adults can easily access web applications, which can collect their conversational speech data automatically, inexpensively and longitudinally. This opens up the possibility of detecting subtle changes in an individual's cognitive abilities over time, to enable early detection of cognitive decline. Such interactive web applications may be able to engage older adults throughout the process of cognitive assessments, e.g. in responding to a series of questions from the CA, or performing a set of requested tasks. In particular, older adults experiencing cognitive decline may need special responses from their interlocutors, such as ``elderspeak” \cite{kemper1998using}. To fulfill this requirement of communicating with older adults effectively, we may reference the inter-personal communication strategies adopted by expert assessors who are well-trained health professionals in engaging participants as they conduct cognitive assessments. Backchanneling is one of such strategies that is proven to be effective \cite{kawahara2016prediction, xiao2013modeling}. Backchanneling refers to verbal responses such as ``uh-huh” or ``hmm”, non-verbal responses such as nodding, smiling and gesturing, as well as {\it both} verbal and non-verbal responses simultaneously given by the listener when his/her counterpart is speaking \cite{poppe2011backchannels, brunner1979smiles, bavelas2011listener, bertrand2007backchannels}. As a form of basic human interaction \cite{heinz2003backchannel}, backchanneling is displayed to show the engagement of the listener and to encourage the speaker to continue speaking without interrupting him/her. The importance of backchanneling has been stated in previous works such as showing understanding and engagement to the speakers \cite{poppe2011backchannels, bavelas2000listeners, heylen2011generating}, and establishing empathy between speakers and listeners \cite{ruede2019yeah}.

In addition, previous work have also discussed the proactive characteristics of backchanneling. It has been shown that comprehension and production can co-exist in conversations \cite{clark2004speaking}. Instead of only being passive recipients of information with reactive backchanneling, listeners can play an active role in a conversation as co-narrators \cite{bavelas2000listeners}. In other words, listeners can give proactive backchanneling such as ``please keep going”. Ortega et al. \cite{ortega2020oh} briefly discussed how to build a proactive listening system using backchannels to influence speakers, which is the first to model backchanneling from a proactive perspective.
% Most previous works of building backchannel algorithms focus on providing generic backchannels which could be used in any kinds of conversation \cite{jain2021exploring, ortega2020oh, ruede2019yeah, schroder2011building}.
%This focus also makes those work hesitate to explore more proactive backchannels from the listeners \xm{don't quite understand this argument...}, given that those proactive backchannels could be specific to the context. As a result,
However, concrete definitions and methodologies of conducting proactive backchannels specific to the conversational context are largely missing in previous studies. To fill in this gap, we define two kinds of backchannels, reactive backchannels (RBCs) and proactive backchannels (PBCs), based on the theories of reactive and proactive backchanneling \cite{ortega2020oh}:

\begin{itemize}
\item Reactive backchannels (RBCs): The listener responds to the previous speaker's utterance directly to show agreement or understanding without intending to take the floor. The response is generic (context-independent) and optional, thereby poses minimal interruption, i.e. the speaker does not need to wait for the response as he/she continues to speak. The response may serve as acknowledgement or assessment. Most RBCs are non-lexical; examples are ``hmm”, ``oh”, ``yeah”.

\item Proactive backchannels (PBCs): The listener responds to the speaker to encourage the speaker to continue speaking without the intent to take the turn. This kind of response is also optional. The response might have more lexical words compared with RBC. Examples are ``please keep going”, ``anything else?”.
\end{itemize}

Based on the definitions of RBCs and PBCs, we aim to investigate their roles in engaging older adults in conversations with CAs to complete NCD screening tasks. Few previous studies have discussed task-based backchanneling, or potential opportunities in adaptive backchanneling depending on tasks and participants. To the best of our knowledge, this study may be one of the earliest to study backchanneling for older adult participants in a cognitive assessment context. Besides, the backchanneling behaviors have been proven to be language-dependent and culture-related \cite{cutrone2014cross, heinz2003backchannel, clancy1996conversational}. To explore the potential for studying backchanneling in low-resource languages, we developed technology to support Cantonese speakers. Cantonese is a predominant Chinese dialect used in Hong Kong, and spoken by over 80 million native speakers worldwide
\cite{eberhard2021ethnologue}. % ~\footnote{[https://www.ethnologue.com/language/yue](https://www.ethnologue.com/language/yue)}
To the best of our knowledge, this is also one of the first efforts that studies backchanneling in Cantonese. This prompts us to ask two research questions (RQs):

\begin{itemize}
\item \RQO
\item \RQT
\end{itemize}

%We hope to extend the previous work by adopting backchannels to a CA conducting Montreal Cognitive Assessment (MoCA) on older adults, where the CA asks the older subject questions for response, encourages the subject to speak more, and collects more data from the subject \xm{this is basically repeating previous messages. You should instead provide a good summary of all the things you have done... }. Our main contributions to the human-computer interaction (HCI) community include:

% \begin{enumerate}
% \item By analyzing of 246 MoCA test conversation between older adults subjects and assessors, we summarize empirical insights of how RBCs and PBCs exist in real world task-oriented conversations, and develop data-driven models to predict RBCs and PBCs;
% \item We build a proof-of-concept system and conduct a between-subjects user study to evaluate our algorithms of predicting RBCs and PBCs in task-oriented conversations. \textcolor{red}{[Todo: concrete finding from user study]}
% \end{enumerate}

\begin{figure*}
\centerline{\includegraphics[width=0.45\linewidth]{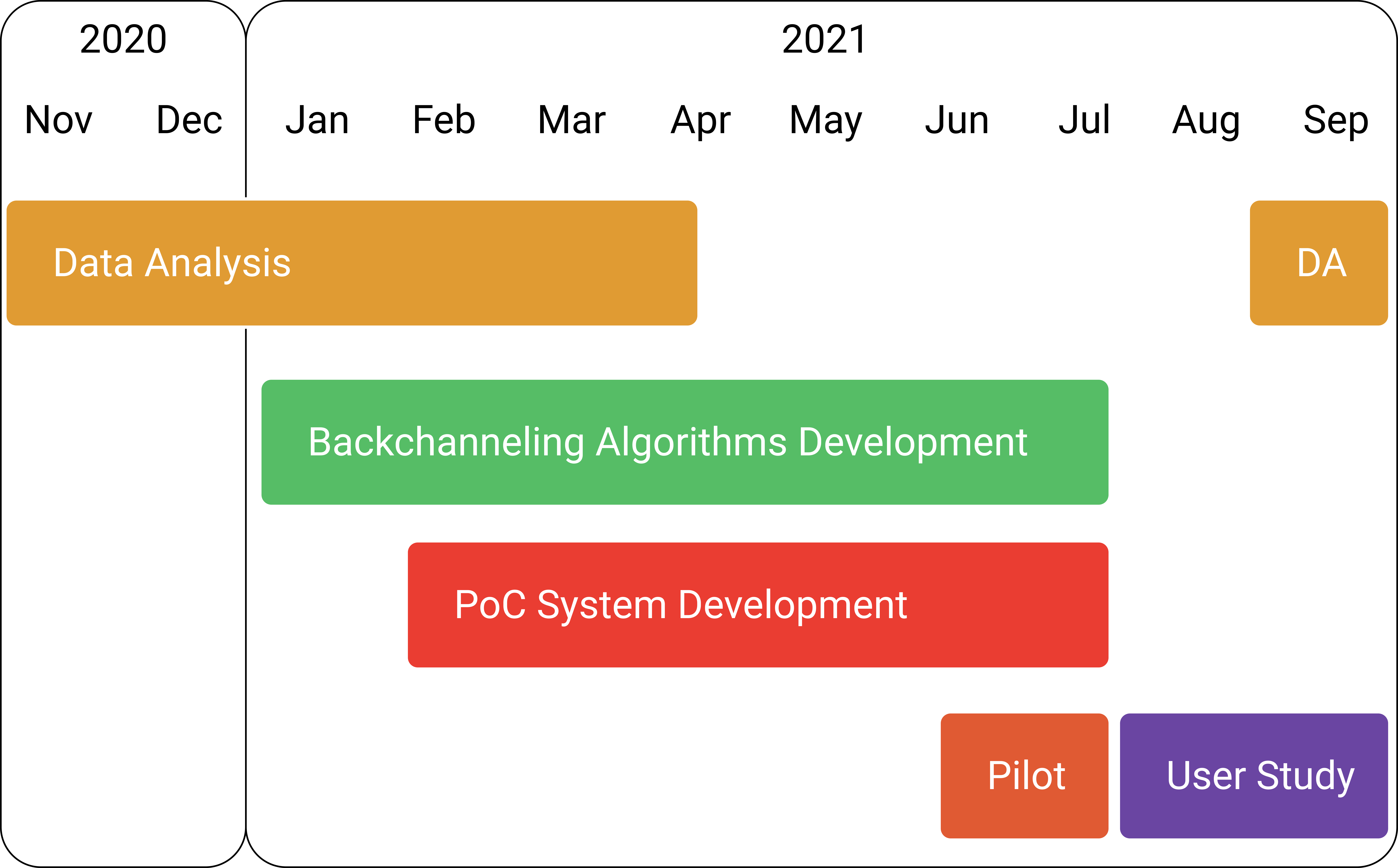}}
  \caption{\todo{A timeline of research activities presented from November 2020 to September 2021.}}
  \Description{The figure shows a timeline of research activities presented from November 2020 to September 2021. Data analysis of existing MoCA dataset was from November 2020 to April 2021; the development of backchanneling algorithm was from January to July 2021; the development of proof-of-concept system was from February to July 2021; the pilot study and user study were from June to July 2021 and July to September 2021 respectively. And the analysis of user study results was from August to September 2021.}
  \label{fig:timeline}
  \vspace{-4mm}
\end{figure*}

To answer these RQs, we first derived empirical patterns of how expert assessors provide RBCs and PBCs in real-world task-oriented NCD screening conversations, by analyzing 246 audio recordings of MoCA test conversations between older adult participants and assessors. Based on the results of this analysis, we developed data-driven ML models to predict proper timing for NCD screening CAs to deliver RBCs and PBCs, trying to mimic human strategies. Next, we built a proof-of-concept, backchanneling-enabled CA system called \textit{TalkTive}, and conducted a between-subjects user study to evaluate older adults' perception of our system in MoCA test conversations, in comparison to a baseline system without backchanneling. \todo{A timeline of research activities presented in this paper is shown in Figure~\ref{fig:timeline}.}

Our main contributions include: 1) developing a backchanneling algorithm to provide RBCs and PBCs with high performance, 2) generating automatic backchannels that offers a positive user experience for older adults, and 3) identifying the type of backchanneling (reactive versus proactive) and discovering that older adults are more receptive to RBCs than PBCs.

%Our main contributions to the human-computer interaction (HCI) community include: 1) our backchanneling algorithm to provide RBCs and PBCs are effective with only 11\% inappropriate backchannels coded by expert assessor, 2) the system-generated backchannels, especially PBCs, are not perceived as disturbing by older adults, and 3) we found that RBCs are received better by older adults than RBCs.

\section{Related Work}

\subsection{Conversational Agents in Healthcare Support for Older Adults}

In face of an ageing global populations, it becomes increasingly important to develop technologies that offer scalable support for older adults, as well as alleviate the pressures of caregiving in the society.  Related research has also attracted growing interests in the HCI community, e.g. \cite{conci2009useful, hanson2010influencing, leung2012older, lindley2009desiring, mitzner2010older, vines2015age}. \todo{One of the main focuses of these HCI studies is to explore potential technological solutions that may reduce the expensive medical costs and enable older adults to manage their own care with greater independence \cite{vines2015age}}, such as providing accessible health monitoring and disease screening solutions \cite{pang2021technology, doyle2015older, doyle2014designing, kononova2019use, lee2020wearable, lewis2017designing}. Among various emerging and evolving technologies, conversational agents (CAs), defined as ``systems that mimic human language and behavior to implement certain tasks for the user via a chat interface, either text-based or voice-based” \cite{abdul2015survey}, has unique advantages in healthcare surveys and disease screening tasks. 
% , which behavior is similar to conversation between patients and clinicians \cite{mathew2019chatbot}. 
Recent research in the adoption and perception of CAs by older adult users show positive results \cite{sengupta2020challenges, kowalski2019older, ziman2018factors, zubatiy2021empowering}. Moreover, it has been proven that patients are more willing to share personal information with virtual therapists than human therapists \cite{lucas2014s}. Applying CAs to support caregiving to older adults also draws recent attention of the CHI community, resulting in a Special Interest Group (SIG) discussion in CHI 2020 \cite{sengupta2020challenges}.

However, adopting CAs to support health monitoring and disease screening is a relatively new trend \cite{kim2019conversational}. For example, while tablet-based applications have been developed to assist clinicians to run tests for neurocognitive disorders (NCD), e.g. MoCA App~\footnote{https://www.MoCAtest.org/app/}, to the best of our knowledge, CAs for NCD tests have not yet been well-studied. Many open problems related to this domain of application exist \cite{kim2019conversational}, e.g., how to make the CAs can ``behave” more like a human as it converses with the user \cite{sengupta2020challenges, jain2021exploring}, as well as how age may impact the user experience in interacting with these CAs \cite{kim2019conversational, vaidyam2019chatbots, pradhan2019phantom}.
%In this new trend there are many angles requiring further study \cite{kim2019conversational}, including how to make the conversation between CAs and people more human-like \cite{sengupta2020challenges, jain2021exploring}, as well as how age might impact the experience of engaging with these CAs \cite{kim2019conversational, vaidyam2019chatbots, pradhan2019phantom}. 

\subsection{Handcrafting Backchanneling Rubrics}

The term ``backchannel” was coined and defined as the messages delivered by listeners in a conversation, without the intent to take a turn \cite{yngve1970getting}. Bangerter and Clark \cite{bangerter2003navigating} defined a backchannel as a listener's response that happens during the speaker's turn, without taking a separate turn.  More recently, researchers categorized backchannels along different dimensions, namely, by content (non-lexical, phrasal or substantive) \cite{iwasaki1997northridge}, by function (backward-looking or forward-looking) \cite{ward2000prosodic}, by relation to the speaker's utterance (generic backchannels/assessments or specific backchannels/continuers) \cite{schegloff1982discourse, goodwin1986between, bavelas2000listeners, stivers2008stance} and by proactivity (reactive or proactive) \cite{tolins2014addressee}.

From the perspective of proactivity, researchers studied the passive characteristics of backchanneling according to the unilateral view of conversations, which may also be referred to as ``reactive tokens” \cite{young2004identifying, clancy1996conversational} or ``response tokens” \cite{gardner2001listeners}. This perspective is known as the reactive backchanneling theory \cite{tolins2014addressee}. On the contrary, other researchers started to study the listener's active participation in the development of a conversation, which also extends to backchanneling \cite{bavelas2000listeners, clark2004speaking, norrick2010listening, norrick2010incorporating, norrick2012listening, beukeboom2009words} -- these studies form the proactive backchanneling theory [86]. We referenced these studies in stating the definition of reactive backchannels (RBC) and proactive backchannels (PBC) in the introductory section. Table \ref{tab:bc_type} shows the connection between proactive perspective and other perspectives.
%From the perspective of proactivity, researchers studied the passive characteristic of backchannels in response to an unilateral view of conversation, which could be reflected by other names referred to backchannels such as ``reactive tokens”  \cite{young2004identifying, clancy1996conversational} and ``response tokens” \cite{gardner2001listeners}. This perspective was named as reactive backchannelling theory \cite{tolins2014addressee}. To the contrary, other researchers started to study the listener's active participation in development of the conversation, which interest then extends to backchannels \cite{bavelas2000listeners, clark2004speaking, norrick2010listening, norrick2010incorporating, norrick2012listening, beukeboom2009words}. Those studies form proactive backchanneling theory \cite{tolins2014addressee}. Based on those studies, we provide the definition of reactive backchannels (RBC) and proactive backchannels (PBC) in Section \ref{sec:intro}. Those definitions from a proactive perspective also have a strong connection with categorizations in other perspectives as shown in Table \ref{tab:bc_type}.

\begin{table*}[]
\begin{tabular}{|l|l|l|l|l|}
\hline
 \textbf{Proactivity} & \textbf{Content} & \textbf{Context} & \textbf{Function} & \textbf{Description} \\ \hline
 Reactive & Non-lexical & Generic & & Vocalic sounds that have little or no referential meaning \\
 & & & Backward- & e.g. ``mm hm”, ``uh huh”, ``yeah” \\
 \cline{1-3} \cline{5-5}
 & Phrasal &  & looking &  Lexical expressions of acknowledgment and assessment \\
 Proactive & & Specific & & e.g. ``really?”, ``I see”, ``I know”, ``good”, ``fine”, ``okay” \\
 \cline{2-2} \cline{4-5}
  & Substantive & & Forward- & Follow-up questions or encouragements to ask the speaker \\
 & & & looking & to talk more e.g. ``Anything else?”,  ``Keep going” \\
 \hline
\end{tabular}
\caption{Comparison among various categorizations of backchanneling: Proactivity, Content, Context (relation to previous speaker's utterance) and Function \todo{\cite{iwasaki1997northridge, ward2000prosodic, schegloff1982discourse, goodwin1986between, bavelas2000listeners, stivers2008stance, tolins2014addressee}}.}
\label{tab:bc_type}
\end{table*}

Research studying human backchanneling behaviors started with handcrafting rubrics for generating backchannels \cite{ward1996using, ward2000prosodic}. Subsequent research referred to the speaker's utterances that trigger backchannels as ``backchannel-inviting cues” \cite{gravano2009backchannel}. Commonly known backchannel-inviting cues are acoustic features, such as pause and pitch (falling or rising slope) \cite{truong2010rule, gravano2009backchannel, ortega2020oh}. There may also be linguistic features such as a final Part-of-Speech bigram in ``DT NN”, ``JJ NN” or ``NN NN” \cite{gravano2009backchannel}. Prosodic and linguistic features are usually combined to achieve a better result if manual transcription or transcriptions from automatic speech recognition (ASR) is available \cite{gravano2009backchannel}. Visual cues such as gaze have also been taken into consideration \cite{poppe2010backchannel}. Research interests in studying user experiences in rule-based backchanneling have expanded from a target audience being adults to specific age groups such as kids \cite{park2017telling}.
%Investigating human backchannel behaviors started from handcrafting rubrics of giving backchannels \cite{ward1996using, ward2000prosodic}. Later research called the speaker's utterances triggering backchannels as ``backchannel-inviting cues” \cite{gravano2009backchannel}. The most important and widely used backchannel-inviting cues to predict backchannels are acoustic features, such as pause and pitch (falling or rising slope) \cite{truong2010rule, gravano2009backchannel, ortega2020oh}. Other features include linguistic features such as a final Part-of-Speech bigram in ``DT NN”, ``JJ NN” or ``NN NN” \cite{gravano2009backchannel}. Prosodic and linguistic features are usually combined to achieve a better result if transcription or automatic speech recognition (ASR) is available \cite{gravano2009backchannel}. Visual cues such as gaze have also been taken into consideration \cite{poppe2010backchannel}. The interest in studying rule-based backchanneling systems is extending to study backchannels occurring in specific age groups such as kids \cite{park2017telling}. However, there is a lack of studies on backchannels targeting older adults. It motivates us to investigate how backchannels are given to older adults, ideally by experts who have been trained to communicate with older adults with patience and empathy \cite{tang2020cccdtd5, ye2021development}.

\subsection{Model-based Backchanneling using Machine Learning Algorithms}
\label{sec:relatedWork3}

Model-based backchanneling approaches divide the task into two parts: prediction and action, referring respectively to the prediction of opportunities for backchanneling from observing the speaker's behaviors and choosing the appropriate type of backchanneling \cite{poppe2011backchannels}. Previous studies formulated the problem in different ways, e.g., focusing on prediction and choosing the same backchannels \cite{jain2021exploring}; bundling multiple binary classifiers to predict different types of backchannels and giving corresponding actions \cite{kawahara2016prediction}; training a multi-class classifier to predict and act at the same time \cite{kawahara2016prediction}. Various machine learning algorithms have been applied, such as locally weighted linear regression \cite{solorio2006prosodic}, Hidden Markov Model (HMM) \cite{morency2008predicting, morency2010probabilistic}, Support Vector Machines \cite{mao2015backchannel}, Long Short-Term Memory networks \cite{ruede2017enhancing, ruede2019yeah, hara2018prediction, jain2021exploring} and hybrid time-delay neural network (TDNN)/HMM system \cite{ortega2020oh}.

Feature engineering is another important component for ML-based methods.  Similar to the acoustic backchannel cues in rubric methods, prosodic features are commonly used as model inputs. Prosodic features include Mel-Frequency Cepstral Coefficients (MFCC), pitch (fundamental frequency), energy, speaking pause (voicing probability) and pitch/power contour \cite{jain2021exploring, ortega2020oh}. Besides, fundamental frequency variation (FFV) , duration,  Spectral Flux, and voice quality-related features could be taken into consideration as well \cite{ruede2019yeah, reidsma2011continuous}. Currently, more comprehensive feature sets are available, such as the ComParE feature set \cite{weninger2013acoustics}, which is a growing set of acoustic features (6,373 in total) \cite{weninger2013acoustics}. ComParE has been widely used in various acoustic recognition tasks, including emotion \cite{schuller2009interspeech, schuller2013interspeech}, speaker \cite{schuller2012interspeech}, eating condition \cite{schuller2015interspeech} and Alzheimer's disease detection \cite{li2021comparative}. This work is the among the first to adopt the ComParE features set for backchanneling prediction.  

%consisting of a series of low-level descriptors (LLDs) and results of LLDs processed by statistical functions 
%It encourages us to apply the ComParE feature set to ML-based methods to achieve a better prediction result.

%As more comprehensive algorithms are applied to backchannel prediction, datasets and feature sets used for training models are also becoming larger. A widely used audio dataset is Switchboard Dialog Act Corpus (SwDA) \cite{ruede2017enhancing, ruede2019yeah, ortega2020oh}, a dialog corpus of 2,438 telephone conversations (260 hours) collected by Godfrey et al. \cite{godfrey1992switchboard} and further annotated by Jurafsky \cite{jurafsky1997switchboard}. The dataset could be task-oriented dialogue \cite{gravano2009backchannel}. Researchers also recorded their own data, such as counseling dialogue \cite{kawahara2016prediction}. At the meantime, latest feature extraction toolkits such as openSMILE are used to derive acoustic features e.g. fundamental frequency variation (FFV), loudness and Mel-frequency cepstral coefficients (MFCCs) \cite{ruede2019yeah, ortega2020oh}.

% In our work, we plan to develop model-based algorithms for RBCs and PBCs respectively through a data-driven method. We investigate an audio dataset with 246 conversations (~30 minutes each, ~120 hours in total) only focusing on the MoCA test, collected by our medical team members.

\section{Data Analysis}
\label{sec:dataAnalysis}

\begin{figure*}
\centerline{\includegraphics[width=400pt]{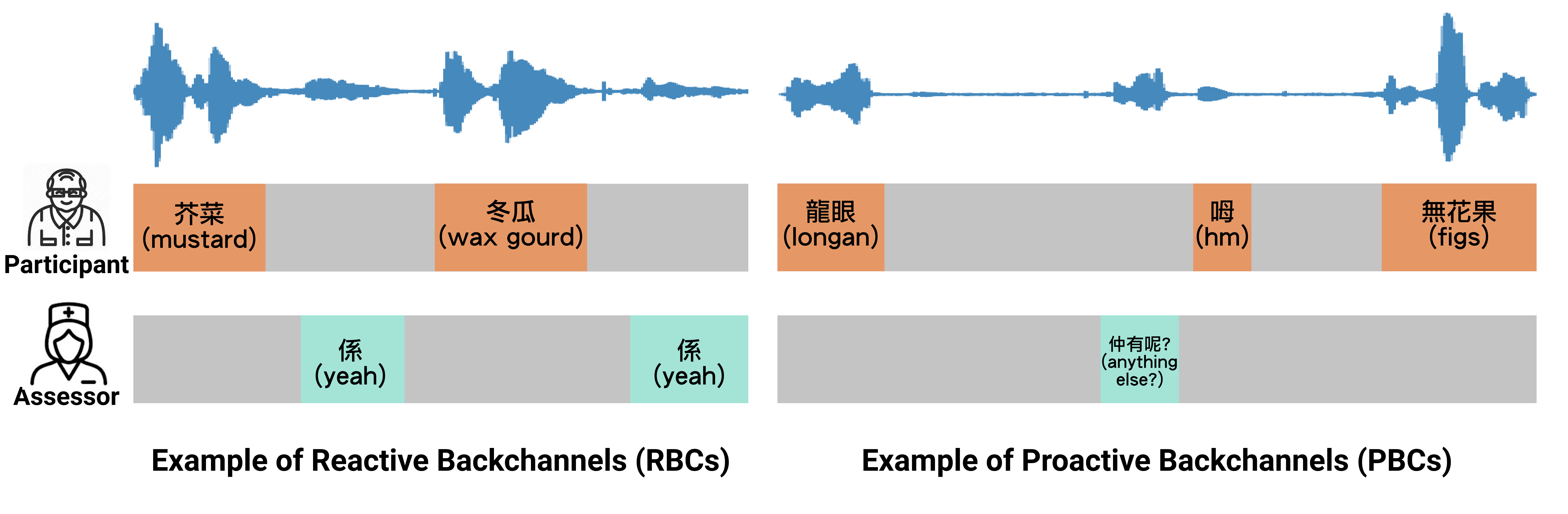}}
  \caption{Examples of reactive backchannels (RBCs) and proactive backchannels (PBCs) from the MoCA dataset. The participant was conducting 1-min verbal fluency task, trying to say as many vegetable/fruit names as possible within one minute and receiving RBCs and PBCs from the assessor.}
  \Description{This figure shows two examples of reactive backchannels (RBCs) and proactive backchannels (PBCs) from the MoCA dataset. The participant was conducting 1-min verbal fluency task, trying to say as many vegetable/fruit names as possible within one minute. The participant received two RBCs after their utterances and one PBC within a silence. }
  \label{fig:bc_example}
\end{figure*}

To address the first research question -- ``\RQO”, we tried to learn from real-world conversations between older adults and trained assessors -- We analyzed a dataset of Montreal Cognitive Assessment (MoCA) recordings based on older adult participants having a conversation with human assessors. This MoCA dataset was collected by 6 experienced clinicians (5 females and 1 male) in 2 years, from June 2015 to July 2017. It included \subjectcount\space Cantonese conversations between trained assessors and older adult participants (171 males and 75 females), each being approximately 30 minutes in duration, and with hand transcriptions aligned at the word-level with speech. The scale of this dataset was comparable with some of the largest datasets used for backchanneling studies, such as SwDA \cite{ruede2017enhancing, ruede2019yeah, ortega2020oh}. Participants in this MoCA dataset were aged between 77 and 94, with an average age of 82.9. Data analyses strictly followed the project's Institutional Review Board (IRB) approval to protect the privacy of participants. \todo{The tasks that participants performed in the MoCA dataset were listed in the Supplementary Material I.}

\subsection{Coding Process for Backchannels}
\label{sec:coding}

To code the assessors' backchannels in this dataset, we followed these three steps:

\begin{enumerate}
\item Inspecting the transcripts and identifying ``backchannel words” related to reactive backchannels (RBC) and proactive backchannels (PBC).  Similar methods of labeling backchannels in the corersponding textual transcripts had been used in previous work \cite{ruede2019yeah},

\item Developing a coding scheme based on RBCs and PBCs, and improving inter-rater reliability (IRR) of that coding scheme, and
\item Building rubrics to auto-coding backchannels on the whole MoCA dataset based on the coding scheme and the ground truth coded by human coders.
\end{enumerate}

\begin{table*}[]
\begin{tabular}{|l|l|l|l|}
\hline
\textbf{Category} & \textbf{Code} & \textbf{Definition} & \textbf{Examples} \\ \hline
RBCs & 1 & Reactive backchanneling to show & oh, hmm, ah \\ & & understanding and agreements & \\ \hline
PBCs & 2 & Proactive backchanneling to encourage & keep going, \\
& & the speaker to speak more &  anything else \\
\hline
\end{tabular}
\caption{Coding scheme based on categorization of content in backchanneling.}
\label{tab:code}
\end{table*}

First, Researcher A with full professional proficiency in Cantonese inspected all the assessors' utterances of 110 conversations, with 15,455 unique utterances in total. The goal was to find ``backchanneling words” which might be candidates of backchannels, such as ``hmm” and ``right”. Researcher A followed the definition of backchannels as ``not occuring in separate turns, but during the speaker's turn” \cite{bangerter2003navigating} to identify utterances that encouraged the participant to talk more without interruption. The inspection resulted in 11 RBC words and 12 PBC phrases (see examples of RBCs and PBCs in Figure~\ref{fig:bc_example}). Then Researcher A developed a coding scheme (see Table \ref{tab:code}) -- If the assessor's utterance was not considered as a backhannel, it would be coded as 0. Researcher A shared the coding scheme with Researcher B, and both started to code the backchannels on the word-level aligned transcripts of four conversations from the MoCA dataset.

Coding 4 conversations were divided into 2 rounds, with 2 conversations in each round. After the first round of coding, Researchers A and B reviewed all the codes and discussed inconsistent codes, finalized the ground truth and refined the coding scheme, and then conducted a second round of coding. Besides the content categorizations of backchannels, rubrics for coding the assessors' backchannels were also developed:

\begin{itemize}
\item (R1) A backchannel should follow an utterance of the participant (which included an utterance followed by a long pause), or a previous backchannel from the assessor.
\item (R2) A backchannel should indicate that the assessor intended to hear from the participant instead of taking a turn turn.
\item (R3) It did not matter whether a backchannel was followed by an utterance (successful backchanneling) or not (unsuccessful backchanneling).
\item (R4) If some information was included in the backchannel, it should be intuitive and minimal given the interaction context, e.g. ``minus”, ``aunt”.
\end{itemize}

\begin{table*}[]
\begin{tabular}{|l|l|l|l|l|}
\hline
\textbf{} & \textbf{\# of Utterances} & \textbf{\# of Backchannels Coded} & \textbf{\# of Consistent Codes (\%)} & \textbf{IRR (Cohen's kappa)} \\ \hline
Round 1 & 588 & A: 46; B: 36 & 563 (95.7\%) & 0.676 \\ \hline
Round 2 & 544 & A: 29; B: 35 & 53 (97.8\%) & 0.803 \\ \hline
\end{tabular}
\caption{Results of two rounds of coding backchannels.}
\label{tab:code_result}
\end{table*}

The results of two rounds of coding are reported in Table \ref{tab:code_result}. The IRR (Cohen's Kappa) between Researchers A and B for the second round of coding was 0.803, which showed substantial agreement between coders \cite{landis1977measurement}. After resolving the mismatches in these two rounds of coding, we obtained the ground truth of the backchannels in those conversations.

Given the large volume of assessors' utterances in the MoCA dataset, we tried to develop a rubric-based method to code the backchannels in \subjectcount\space conversations automatically. Below are the rubrics we used to auto-code RBCs and PBCs.

Rubrics to auto-code RBC:
\begin{itemize}
\item (RBC-R1) only one word in the RBC words, e.g. ``hmm”, ``oh”, ``uh”, ``ah”, ``huh”;
\item (RBC-R2) at least 1000ms after the assessor's previous utterance and 1000ms before assessor's next utterance.
\end{itemize}

Rubrics to auto-code PBC: \begin{itemize}
\item (PBC-R1) at most 8 words (the longest PBC phrase has 3 words), including PBC phrases e.g. ``keep going”, ``next”, ``understand”, ``great”, ``awesome”, ``no rush”, ``anything else”;
\item (PBC-R2) at least 1000ms after the assessor's previous utterance and 1000ms before assessor's next utterance.
\end{itemize}

The lists of RBC words and PBC phrases with English translations  provided in Supplementary Material II. The reason for adding RBC-R2 and PBC-R2 was to guarantee that the utterance was not a part of a longer utterance such as ``Hmm...time is up”, which might involve taking a turn. Our auto-coding scheme achieved substantial agreement with the ground truth (\textit{Cohen's kappa = 0.774}) \cite{landis1977measurement}.  We used these rubrics to code all the assessors' utterances in the MoCA dataset, resulting in 2,732 RBCs and 2,037 PBCs.

\subsection{Insights from Data Analysis}
\label{sec:DataAnalysisInsight}
\begin{figure*}
\centerline{\includegraphics[width=350pt]{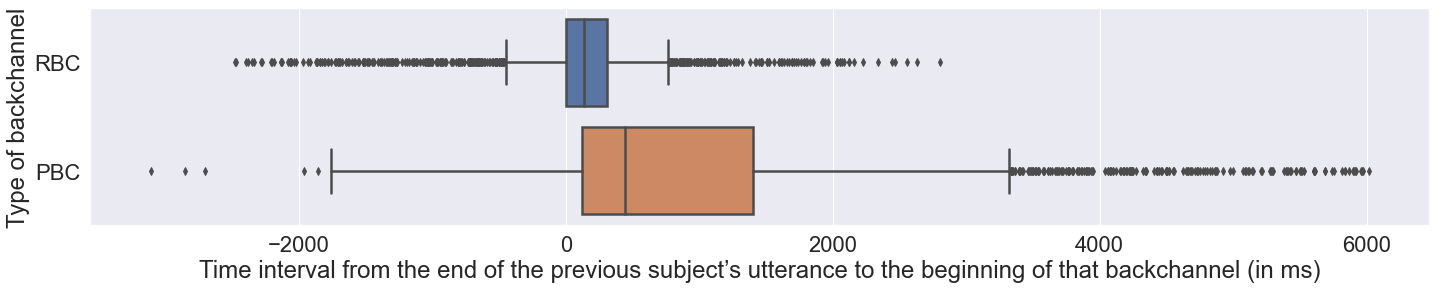}}
  \caption{The assessor's PBCs may begin after a longer time interval from the end of the previous participant's utterance, as compared with the assessor's RBCs. The time interval may also be negative, which implies an overlap with previous participant's utterance.}
  \Description{This figure shows two boxplots of the distribution of interval from the end of the previous participant's utterance to RBCs and PBCs. Those boxplots indicate that the assessor's PBCs may begin after a longer time interval from the end of the previous participant's utterance, as compared with the assessor's RBCs.}
  \label{fig:ds_result}
  \vspace{-4mm}
\end{figure*}

As stated in RQ1, we aimed to study the timing of backchanneling in task-oriented conversations. We planned to analyze coded backchannel occurrences to gain insights from real assessors' behaviors. To understand the timing of RBCs and PBCs after participants' speech utterances, we first drew a boxplot to show the distribution of time intervals between the end of participants' previous utterance and the beginning of backchannel (see Figure~\ref{fig:ds_result}).

In Figure~\ref{fig:ds_result} we observed that assessor's RBCs tended to have shorter time interval and much smaller variance than those of the PBCs ($t=26.18$, $p<<0.01$, after removing outliers over 2 SDs).  This indicated that that RBCs were likely to occur after a certain duration of pausing that followed the participants' speech.  In comparison, PBCs were likely to occur significantly later, and hence they did not serve as an immediate response to the speaker utterance. This observation aligned with our definition of RBCs -- that they were generic and independent of other factors such as contextual information (e.g., the task, participant, etc.), and the timing of RBCs after the pauses were relatively stable ($M=178.2, SD=1335.7$). As for PBCs, their timing after the pauses were much longer with a larger variance ($M=1285.1, SD=2371.9$), which might reveal that there were factors other than pause duration that affected the timing of PBCs. Hence, we inspected the distribution of PBCs according to three other factors: within-task progress, between-task differences, and between-participant differences, and summarized our observations according to possible elements that might affect how proactive the human assessors may give PBCs.
%factors, progress of task, type of tasks and characteristics of subjects that might affect the proactive level of human assessors to give PBCs.

\begin{figure*}
     \centering
     \begin{subfigure}[b]{0.33\textwidth}
         \centering
         \includegraphics[width=1\textwidth]{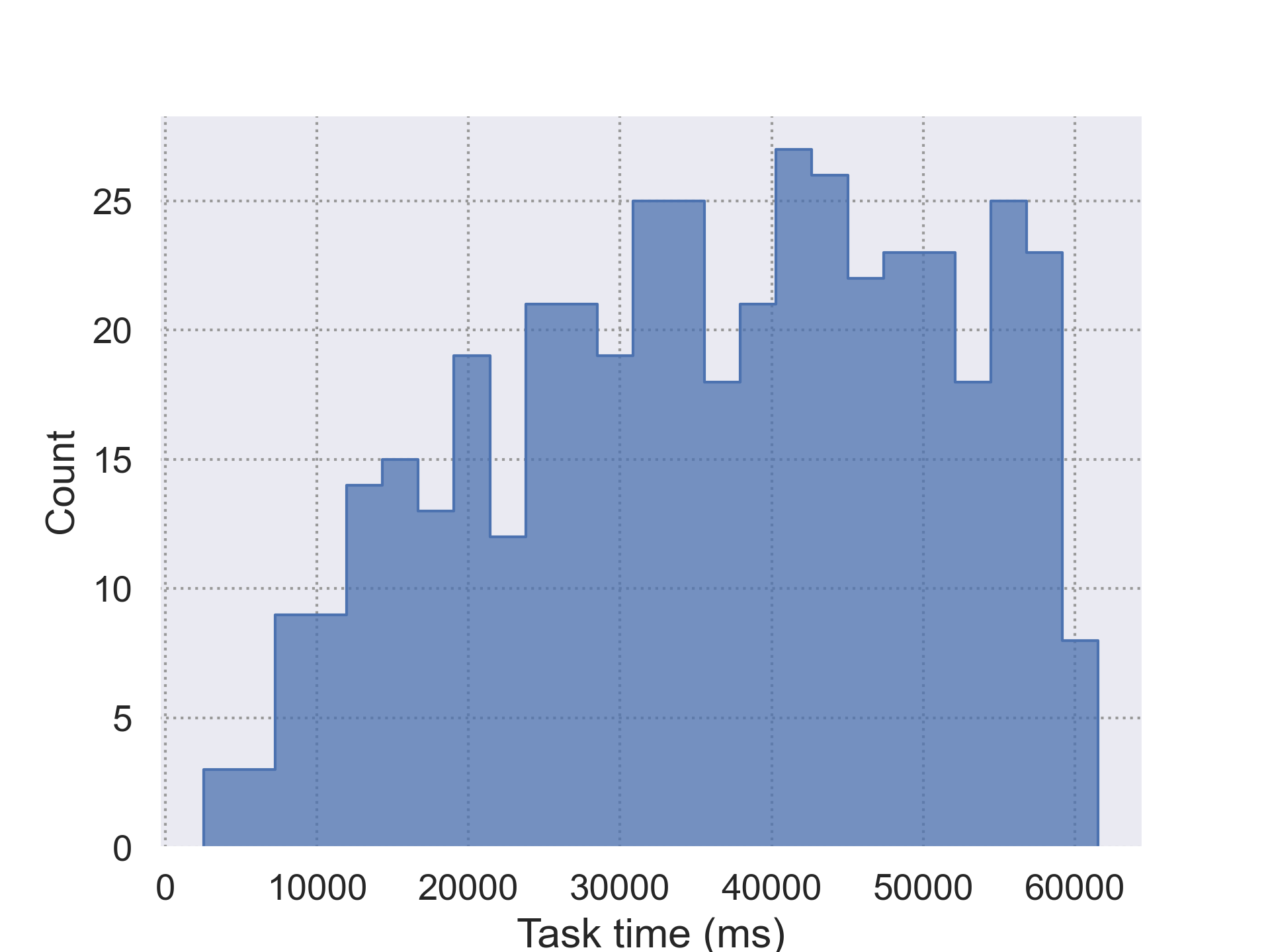}
         \caption{}
         \label{fig:HistProgress}
     \end{subfigure}
     \begin{subfigure}[b]{0.33\textwidth}
         \centering
         \includegraphics[width=1\textwidth]{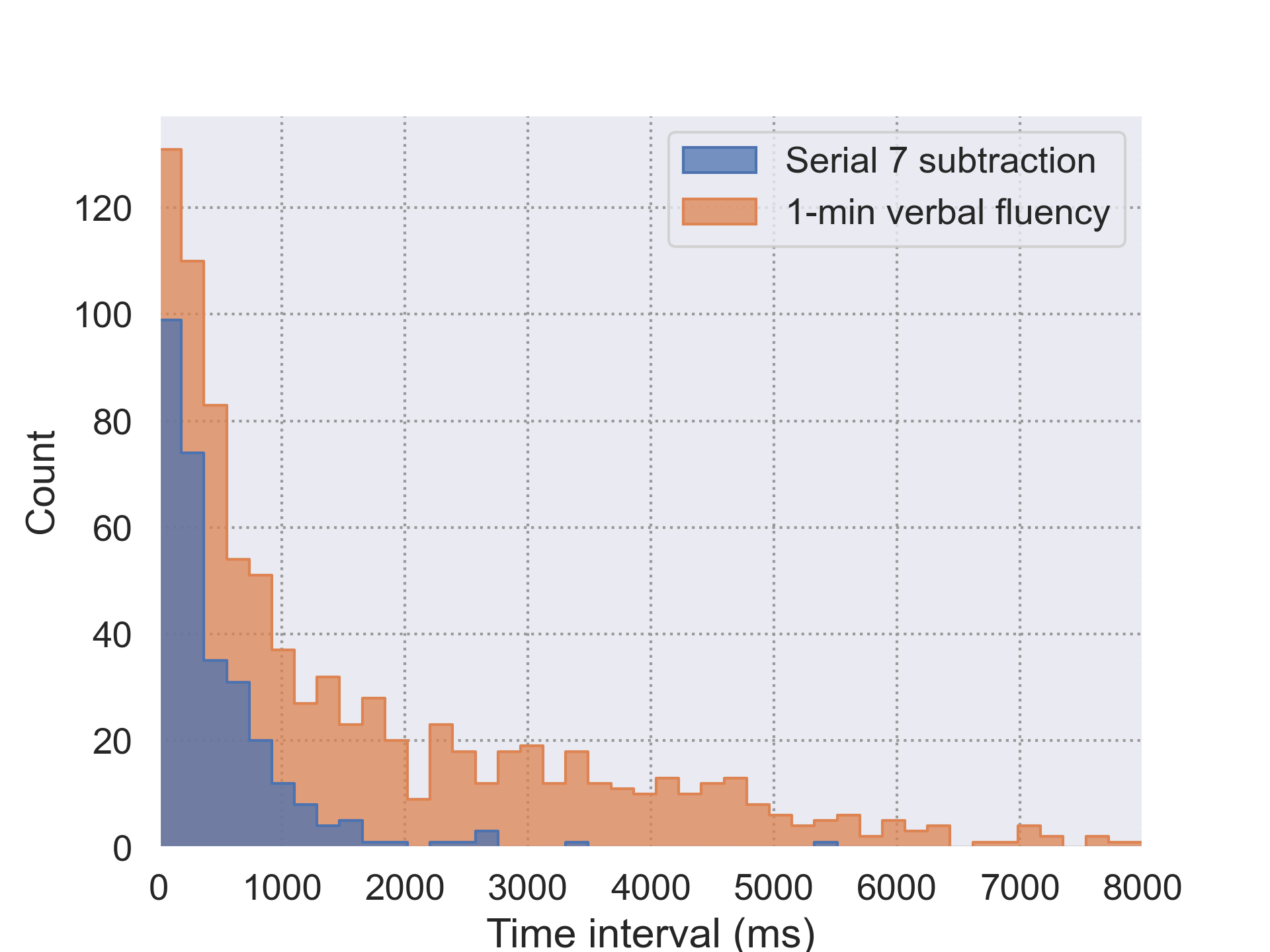}
         \caption{}
         \label{fig:HistPause}
     \end{subfigure}
     \begin{subfigure}[b]{0.33\textwidth}
         \centering
         \includegraphics[width=1\textwidth]{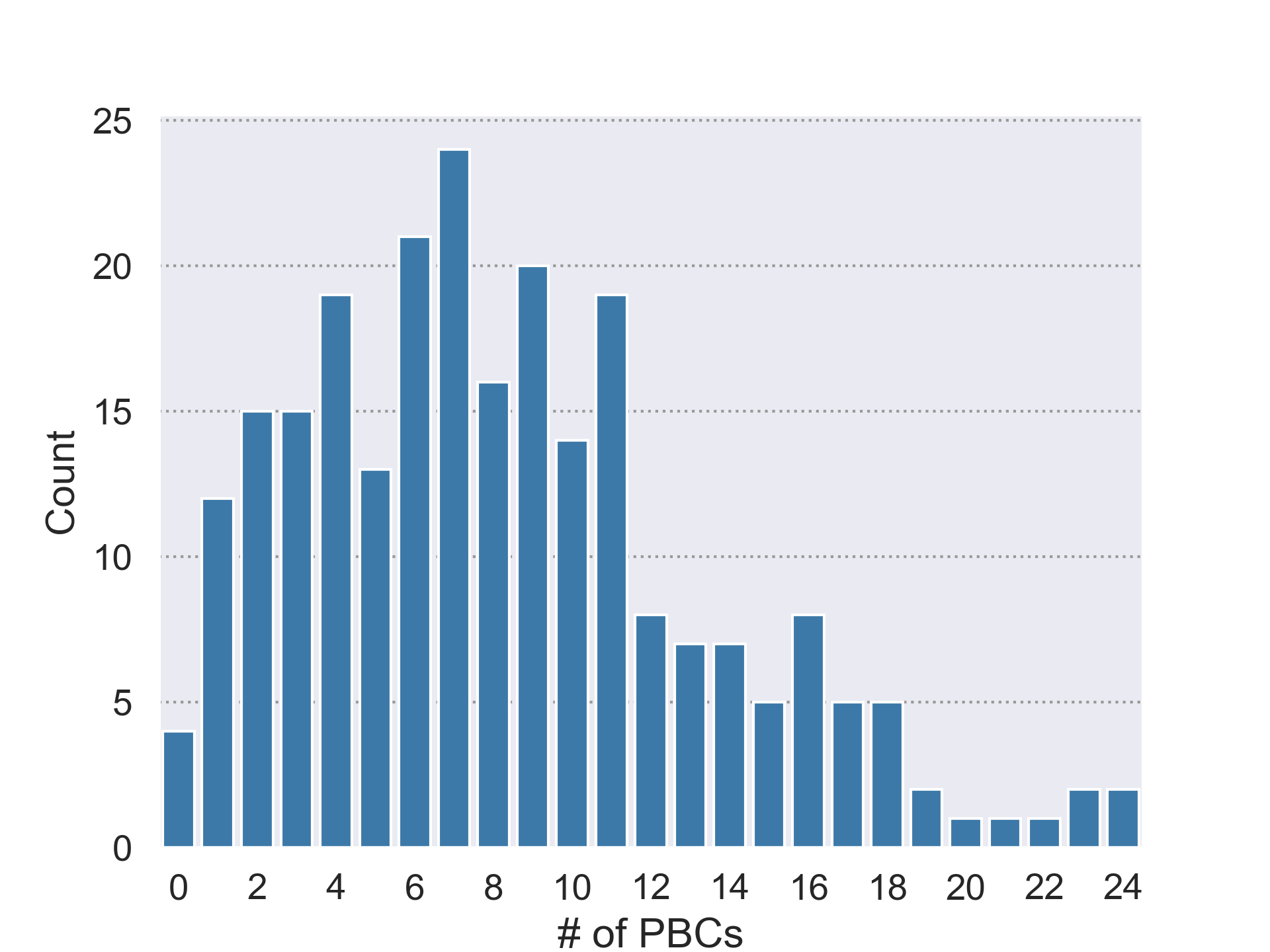}
         \caption{}
         \label{fig:HistParticip}
     \end{subfigure}
        \caption{The number PBCs given by expert assessors varied along 3 dimensions: within task (a), between tasks (b) and between subjects (c). Plot (a) describes the distribution of PBCs throughout the progress of the 1-min verbal fluency tasks,  Plot (b) describes the distribution of time intervals and Plot (c) describes the distribution of the number of PBCs received by participants.}
        \Description{This figure contains three plots of the PBC distributions along 3 dimensions: within task, between tasks and between subjects. Plot (a) shows a skewed normal distribution describing the distribution of PBCs throughout the progress of the 1-min verbal fluency tasks, which peaks in the second half of the task. Plot (b) shows two long tails describing the distributions of time intervals between the speaker's utterance and the PBC in two kinds of tasks, serial 7 subtraction and 1-min verbal fluency. The intervals before PBCs in 1-min verbal fluency tended to be longer than those in serial 7 subtraction. Plot (c) shows a skewed normal distribution of the number of PBCs received by participants, which peaks at 7.}
        \label{fig:dataHists}
\end{figure*}

\subsubsection{Within-task differences in PBCs given by expert assessors}
\label{sec:dataProgress}
We first investigated the timing of PBCs as a task progresses -- Figure~\ref{fig:HistProgress} plotted the distributions of PBCs among all the 1-min verbal fluency tasks.  We observed that assessors gave more PBCs halfway into the task, and then the number fell before the task ended. One potential reason may be that the participant ran out of answers later in the task and assessor tried to use more PBCs for encouragement, while the assessor also noted the time limit and tended to give fewer PBCs when there were only a few seconds remaining.  These results showed that assessors may vary their proactive levels dynamically according to task progression.

%% between task analysis (factor: tasks)
\subsubsection{Between-task differences in PBCs given by expert assessors}
\label{sec:dataTask}
We noticed the differences in the PBCs given across tasks in the MoCA tests.  Considering the number of PBCs given, we identified three major types of tasks: Type I) \textit{tasks requiring a one-off response}, where no PBCs were given, e.g., ``please repeat the sentence…”; Type II) \textit{tasks requiring a series of responses in a given time}, where several PBCs may be given, such as the 1-min verbal fluency task ``please say as many animal names as possible in one minute”, and the serial 7 subtraction task ``please use 100 minus 7, and continue to subtract 7”; and Type III) \textit{tasks requiring open-ended self-disclosure}, where few PBCs may be given, e.g., ``where is your favorite place and why”.

Even for the same type of tasks, such as the 1-min verbal fluency task and the serial 7 subtraction task, both of which required a series of responses from participants, there existed differences in the duration of tolerated silence before a PBC was given (see Figure~\ref{fig:HistPause}). Most PBCs for the serial 7 subtraction task occurred within 1 second after participants' previous answer, while for the 1-min verbal fluency task, the corresponding interval could be as long as 10 seconds. A possible explanation may be due to the level of difficulty of the task. For MoCA, many older adult participants stated that the serial 7 subtraction task was quite difficult for them. The expert assessors were mindful of the level of task difficulty and tended to give more PBCs to encourage the participant for more difficult tasks.

\subsubsection{Between-participant differences in PBCs given by expert assessors}
\label{sec:dataParticipants}
%Towards developing a backchanneling strategy adaptive to participants' characteristics, we investigated between-participants difference of receiving PBCs. Because participants had intrinsic differences, the support needed from assessors to facilitate the participant to complete the test,
Figure~\ref{fig:HistParticip} shows the total number of PBCs received by each participant in the MoCA dataset. Although each participant was asked to conduct the same set of tasks, the counts of PBCs received per person varied substantially from 0 to 24. This result suggested that there existed between-participants difference in how expert assessors gave PBCs.

% This is reasonable for a task-oriented conversation, as the individual differences in testing ability would affect how much support we should provide to them.
% Based on the observations, assessors in have the ability to adapt their encourage behaviors towards such individual difference. 

%\subsubsection{(Motivation for pause score)}
%\label{sec:dataPauseScore}

% \textcolor{blue}{We firstly look into the utterance level of backchannel locations. Among 2,732 assessor's RBCs, 683 (25\%) of them have an overlap with previous participant's utterances and 2,049 (75\%) of them happen after previous utterances. The counts for PBC become 264 (13\%) and 1,773 (87\%) for overlap and non-overlap PBCs respectively. It shows that PBCs tend to wait for a full stop of previous utterances to be triggered.} We are interested in the exact location where backchannels are placed.

\section{Methodology of Backchanneling}
\label{sec:methodology}

\begin{figure*}
\centerline{\includegraphics[width=0.8\linewidth]{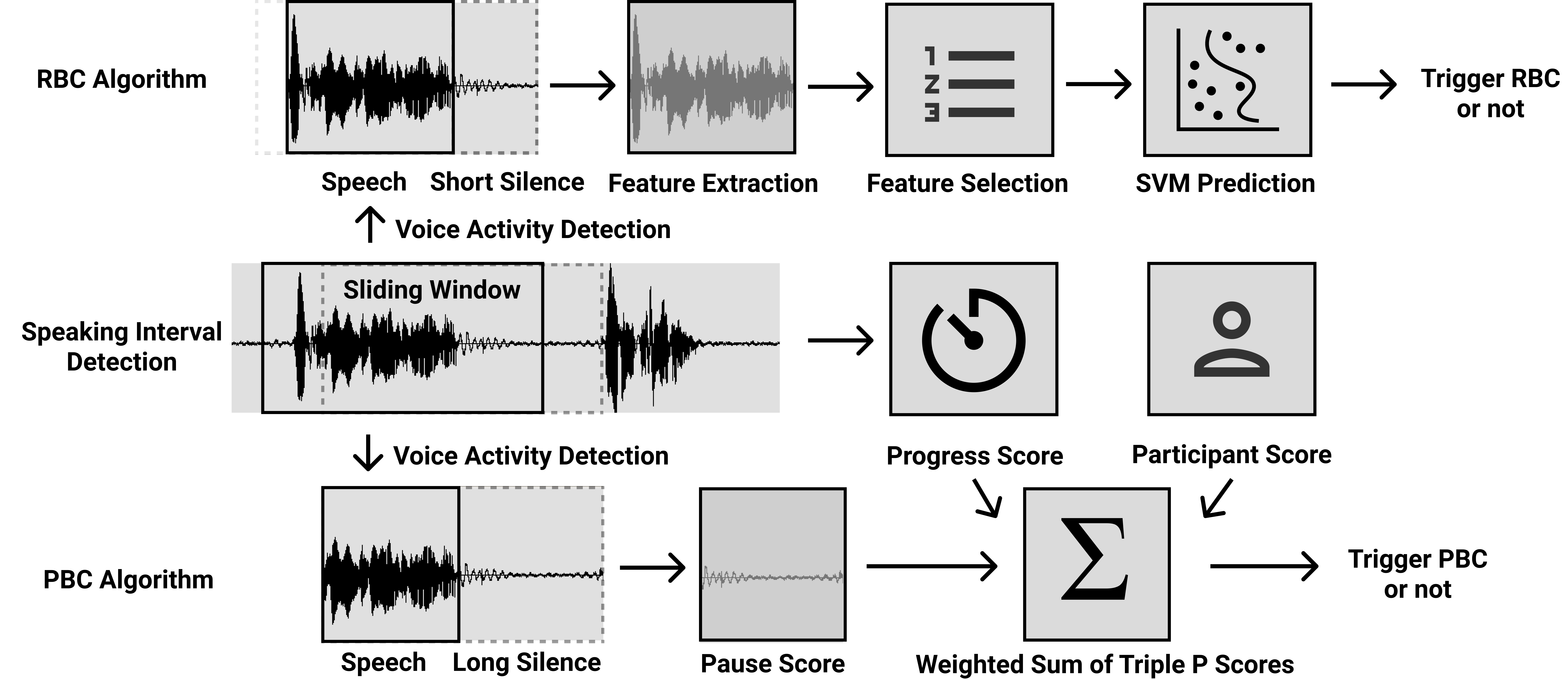}}
  \caption{Overview of the pipeline for processing speech data and generating RBC and PBC decisions.}
  \Description{This figure shows an overview of the pipeline for processing speech data and generating RBC and PBC decisions. The description could be found in the first paragraph of section methodology of backchanneling.}
  \label{fig:pipeline}
\end{figure*}

Given the difference in functions and timing between reactive backchannels (RBCs) and proactive backchannels (PBCs), we developed two different models to generate these two kinds of backchannels. This section describes the models and discusses how they are integrated into a functional backchanneling system. To give an overview, when a speaking interval was detected, the previous speaking utterance followed by this interval would be segmented. Then, the speaking utterance was fed into OpenSmile feature extractor to obtain selected ComParE features, and those features were used by the trained SVM model to make RBC predictions. For PBC prediction, three scores were independently calculated and a weighted sum of those scores was used to trigger PBCs. Participant Score was calculated by passing the speech of the exact same sentences to SVM model. In a certain task, Progress Score was updated as the task was going on, and when speaking interval occurs, Pause Score would be calculated as the length of pauses at that time stamp. The pipeline of processing speech data is illustrated in Figure \ref{fig:pipeline}.

\subsection{Reactive Backchanneling Algorithm}
Since RBCs usually follow the previous speaker's utterances as a direct signal of understanding or agreement \cite{tolins2014addressee}, we pose the problem of finding RBCs as a binary classification problem -- Given a speaker's utterance, the RBC prediction model will make decision regarding whether the utterance contains the backchanneling acoustic cues \cite{ortega2020oh} to trigger an RBC.
Building on this definition, we investigated and compared multiple features and models, then proposed a method leveraging the ComParE feature set, a LASSO-based feature selection algorithm, and SVM classifier to predict RBCs.

\subsubsection{Feature Engineering}
%
%The selection of features plays an important role for Backchannel task, especially when data volume is limited, it is difficult for models to find hidden decision cue in data distribution. 
%In hence, discrimination of input features is rather important for model performance.
%Many papers have investigated on acoustic backchannel features, it mainly involves pitch level [ref], pitch/power contour [ref], speaking pauses [ref] and prosodics features [ref]. 
%Even though, most of relevant works focus on English context. 
%According to researches from Xia Mao et.al [ref], backchannel cue in Standard Mandarin should be specific due to culture difference and different tone changes in language level.
%In our system, task focus Cantonese conversation, which has vast pitch and tone difference with Mandarin [ref], it's risky to directly borrow feature strategy from other language. 
%For this reason, a novel feature selection method was introduced to further investigate on Cantonese backchannel features,  involving ComParE feature set and LASSO-based feature selection algorithm.

Instead of using existing features from other languages and dialects, we performed feature engineering to retrieve appropriate features from Cantonese utterances, since  backchanneling features are largely affected by languages and cultures \cite{cutrone2014cross, heinz2003backchannel, clancy1996conversational}. As introduced in \ref{sec:relatedWork3}, dominant acoustic features are extracted based on the ComParE feature set \todo{\cite{schuller2013interspeech, schuller2015interspeech}}, for triggering backchanneling, and all the spoken interactions were in Cantonese.

A LASSO-based algorithm was used to perform feature selection and reduce feature dimensions. 
The algorithm utilized L1 penalty on linear regression models to force the model to have a sparse weights distribution, by which the features with non-zero weights are selected. 
We noticed that the selection process was sensitive to the random initialization parameters. Hence, we further applied a stability selection algorithm \cite{meinshausen2010stability, shah2013variable}: instead of using the result of one round of selection on the whole dataset, we used stability selection to randomly subsample half of the data for N rounds, aggregated all the features selected and recorded the times they get selected. Then a threshold was used to keep the most frequently selected features. \todo{In our experiments, this threshold was set to $0.6$ as default.} As a result, 34 out of 6373 features were selected as input features to train the RBC prediction model. In addition, we also used a prosodic feature set as a baseline, including the 13-dimensional Mel-Frequency Cepstral Coefficients (MFCC), fundamental frequency (F0), and sum square energy, which were commonly used in many previous works \cite{macneil2019ineqdetect, jain2021exploring, ortega2020oh}. Selected ComParE features and prosodic features were listed in the Supplementary Material III.

%The results of our feature selection aligns with previous work on acoustic features for backchannel prediction, with a simliar focus on prosodic feastures such as 13-dimentinal Mel-Frequency Cepstral Coefficients (MFCC), fundamental frequency (F0), and sum square energy [ref]. While leveraging the ComParE feature set enables us to gain an in-depth view of subtle features for backchannel prediction.

\subsubsection{Theoretical Experiments for Reactive Backchanneling}
\label{section:Models and Experiments}
A series of experiments were conducted to obtain the best-performing configuration of features and models.
We evaluated the selected ComParE features using the Multilayer Perceptron (MLP) and Support Vector Machine (SVM) models. A reference baseline experiment was conducted using prosodic features and the Long Short-Term Memory (LSTM) model, according to the state of the art \cite{jain2021exploring, ortega2020oh}. The reason that we did not combine the ComParE feature set with the LSTM model is that ComParE feature set did not include the time-domain information of speech segments, and hence the LSTM model cannot be used.
%data in a certain segment, so it could not be applied to sequence models such as LSTM.

As for the training data, 2,732 RBCs coded in \ref{sec:coding} were used. According to previous work \cite{gravano2009backchannel}, we considered the participants' utterances before coding the assessors' backchannels as \textit{RBC cues}, which may be more likely to contain acoustic information for triggering RBCs
%as compared with  than other subjects' utterances. 
The negative samples were randomly selected from the participants' utterances that were not followed by assessors' backchannels, referred to as \textit{non RBC cues}. 
For a legitimate comparison, the sampling process took both participant and task distributions into consideration, i.e. if there were \textit{n} RBC cues coming from the participant \textit{x} in task \textit{t}, then \textit{n} non-RBC cues would be randomly sampled from the participant \textit{x} in task \textit{t}. In this way, we obtained a balanced data of RBC cues and non RBC cues.
% distribution are speaker-independent and task-independent, expecting the model emphasizes on backchannel-related information.

\begin{table*}[]
\begin{tabular}{llllll}
\hline
\textbf{Feature}                   & \textbf{Model} & \multicolumn{1}{c}{\textbf{Accuracy}} & \multicolumn{1}{c}{\textbf{Precise}} & \multicolumn{1}{c}{\textbf{Recall}} & \multicolumn{1}{c}{\textbf{F1}} \\
\hline
\multirow{1}{*}{Prosodic} & LSTM  & 0.615/0.638 & 0.573/0.610 & \textbf{0.866/0.859} & 0.689/\textbf{0.714} \\
 \hline            
\multirow{2}{*}{ComParE}  & MLP   & 0.650/0.642 & \textbf{0.670/0.671} & 0.579/0.623 & 0.615/0.646\\
                          & SVM   & \textbf{0.692/0.655} & 0.656/0.646 & 0.793/0.760 & \textbf{0.718}/0.695\\
\hline
\end{tabular}
\caption{Experiment results on RBC prediction models. Metrics are denoted as Cross Validation/Test.}
\label{tab:RBCModel}
% \centering
\end{table*}

From the experimental results (see Table \ref{tab:RBCModel}), we observe that the results based on selected ComParE features and SVM generally outperformed the LSTM baseline, while the baseline method had a higher recall rate. As stated in the definitions of RBCs and PBCs, backchanneling is an optional behavior and it is not necessary to give a backchannel within a certain timing, so the recall metric may not be a major concern this context. Therefore, taking performance and inference speed into consideration, the SVM model with the ComParE feature set is selected to be the best one for implementation in our system.

\subsection{Proactive Backchanneling Algorithm}
The function of PBCs is to encourage speakers to continue talking, and the most intuitive way of triggering PBCs is to take place after the users have stopped speaking for a while.  More specifically,  
%PBCs when users stopped answering questions for a while. The fact that 
PBCs tend to be triggered by long pauses instead of speech segments and short pauses.  This implies that the methodology for predicting RBCs may not be applicable to predicting PBCs. Moreover, as discussed in \ref{sec:DataAnalysisInsight}, trained assessors were observed to take task progress, type of tasks, and characteristics of participants into consideration when giving PBCs.  Hence, we introduce a comprehensive scoring method, \textit{Triple P Scoring Method}, to imitate those adaptive PBC strategies. The \textit{Triple P Scoring Method} included three main components: \textit{\textbf{P}ause Score}, \textit{\textbf{P}rogress Score} and \textit{\textbf{P}articipant Score}. We calculated the final PBC score through a weighted sum of those three scores, and a PBC will be triggered if the PBC score exceeds  a threshold. The threshold was determined by the data collected through piloting our system with 10 participants. Below we introduce how we calculated those three scores for Type II tasks (e.g. 1-min verbal fluency task and serial 7 subtraction task), where most PBCs occurred in the MoCA dataset.

\subsubsection{Pause Score}

As mentioned earlier, pausing is the most intuitive cue for providing PBC, so the Pause Score is a critical component for triggering PBC.  

\begin{figure}
     \centering
     \begin{subfigure}[b]{0.4\textwidth}
         \centering
         \includegraphics[width=1\textwidth]{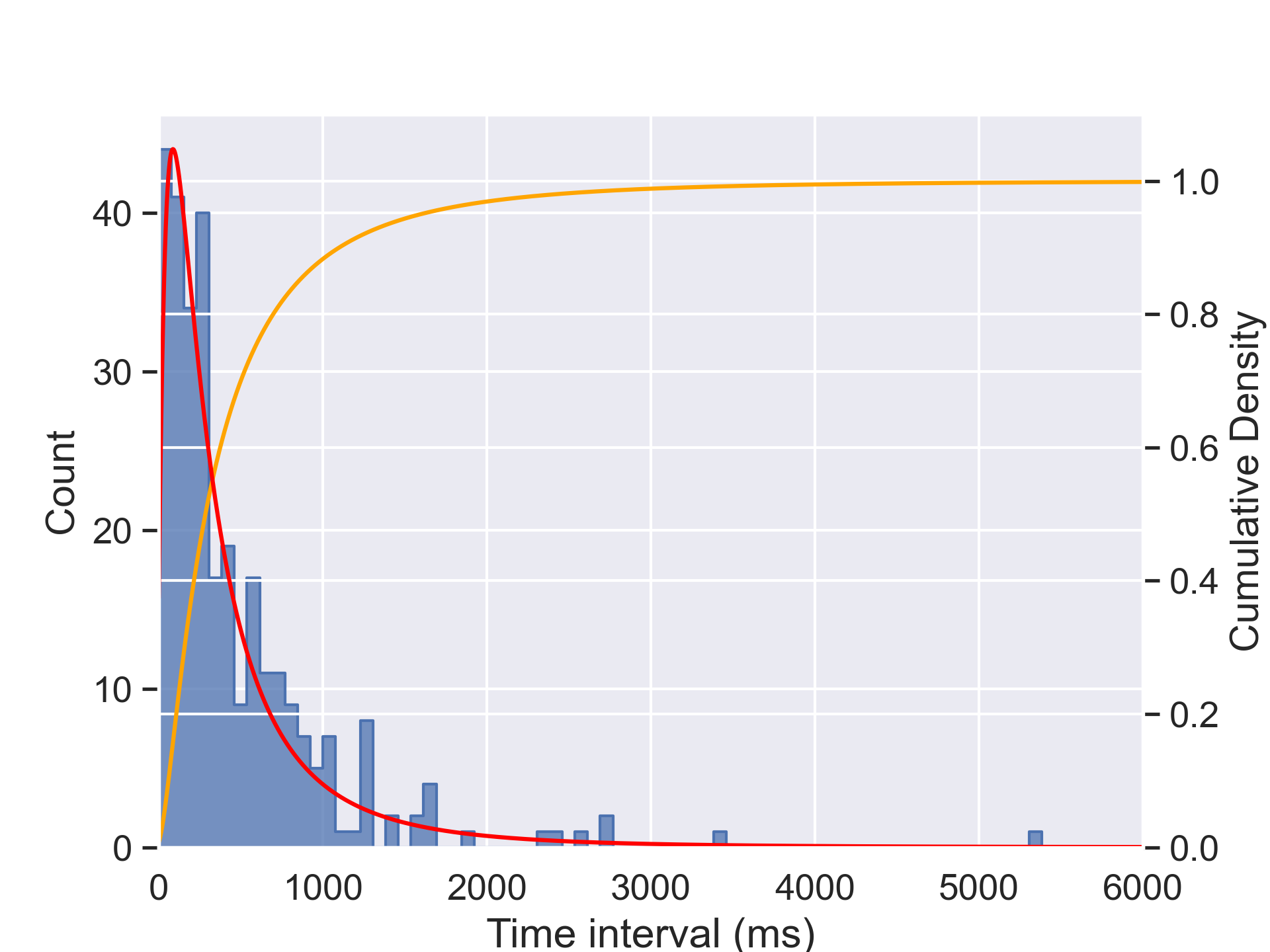}
         \caption{}
         \label{fig:pauseTaskA}
     \end{subfigure}
     \begin{subfigure}[b]{0.4\textwidth}
         \centering
         \includegraphics[width=1\textwidth]{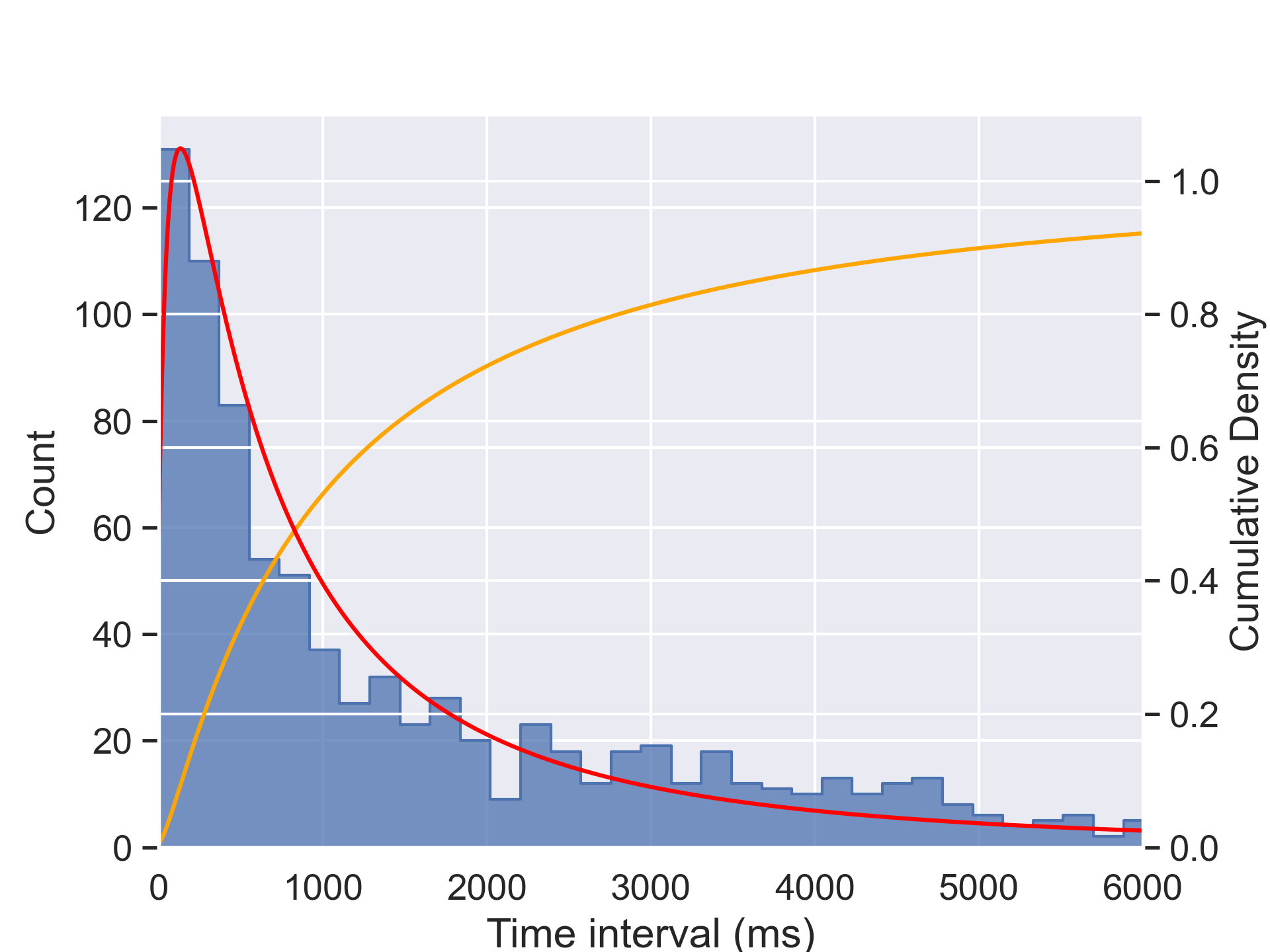}
         \caption{}
         \label{fig:pauseTaskB}
     \end{subfigure}
        \caption{Distributions of pauses before PBCs show a difference between two tasks: a) serial 7 subtraction task b) 1-min verbal fluency task, where assessors might have a lower tolerance of silence and were more inclined to give PBCs faster in a) compared with b). Probability Density Function is in red and Cumulative Density Function is in orange.}
        \Description{This figure shows a difference between the pauses before PBCs in two tasks: serial 7 subtraction task and 1-min verbal fluency task, where assessors might have a lower tolerance of silence and were more inclined to give PBCs faster in serial 7 subtraction task compared with 1-min verbal fluency task.}
        \label{fig:pauseFitting}
\end{figure}

First, a Log-normal distribution was selected through distribution selection with minimum sum square error in the MoCA dataset, as a Probability Density Function (PDF) to model the distribution of the intervals between PBCs and their related speaker utterances (see Figure \ref{fig:pauseFitting}). Then, we calculated the Cumulative Density Function (CDF) of the modeled PDF, and the value of CDF at a certain point of time was used as the Pause Score of that moment.
According to the CDF, the obtained Pause Score was in the range of (0, 1) and increases as the participant's pause becomes longer.

The distribution of the interval between participant's utterances and PBCs is:

\begin{equation}
\label{eq:lognorm}
PDF_{lognorm}(t_{pau};z,s) = \frac{1}{sz\sqrt{2\pi}}exp(-\frac{log^2(z)}{2s^2})
\end{equation}

\begin{equation}
\label{eq:lognorm_z}
z=\frac{t_{pau}-\mu}{\sigma}
\end{equation}

where $t_{pau}$ means the silence time of participant, and $\mu$, $\sigma$ and $s$ are parameters of log-norm distribution estimated by maximum likelihood from the MoCA dataset.  The Pause Score could be expressed as:

\begin{equation}
\label{eq:PauseScore}
Score_{pau}=CDF_{lognorm}(t_{pau};z,s)=\int{PDF_{lognorm}(t_{pau};z,s)} dx
\end{equation}

Besides, as shown in \ref{sec:dataTask}, the s of the intervals between the speaker's utterances and the PBCs are largely depending on tasks.
For example, it seems that assessors tended to have lower tolerance of silence and gave more encouragement to the older adult participants whey they were undertaking relatively harder tasks, e.g. serial 7 subtraction. Hence, we differentiated CDFs and PDFs for different tasks. For instance, Figure \ref{fig:pauseFitting} compares two tasks: the serial 7 subtraction task and 1-min verbal fluency task. We can see that the Pause Score of the serial 7 subtraction increases more rapidly than that of the 1-min verbal fluency task, which means that algorithm tended to provide more PBCs in serial 7 subtraction than in 1-min verbal fluency task.

\subsubsection{Progress Score}
As discussed in \ref{sec:dataProgress}, there was a clear difference in the number of PBCs provided by the assessors as the task progresses. For example, the assessor would provide more PBCs during the second half of the one-minute naming task.  We model this behavior of the assessors with the Progress Score.
Similar to Pause Score, a PDF obtained from distribution selection was used to model the distribution of PBCs during the task period. Moreover, to derive a computable score, we discretized the PDF into Probability Mass Function (PMF) in bins of $100 ms$, and then used a scaling factor to rescale the maximum of PMF to 1, by which we obtained a score in [0, 1]. 
Based on the MoCA dataset, the Skew-Normal distribution was selected as PDF, and this process can be described by the following function:

\begin{equation}
\label{eq:ProgressScore}
Score_{pg}=k*PMF_{skewnorm}(t_{task})
\end{equation}

\begin{figure}
\centerline{\includegraphics[width=0.4\textwidth]{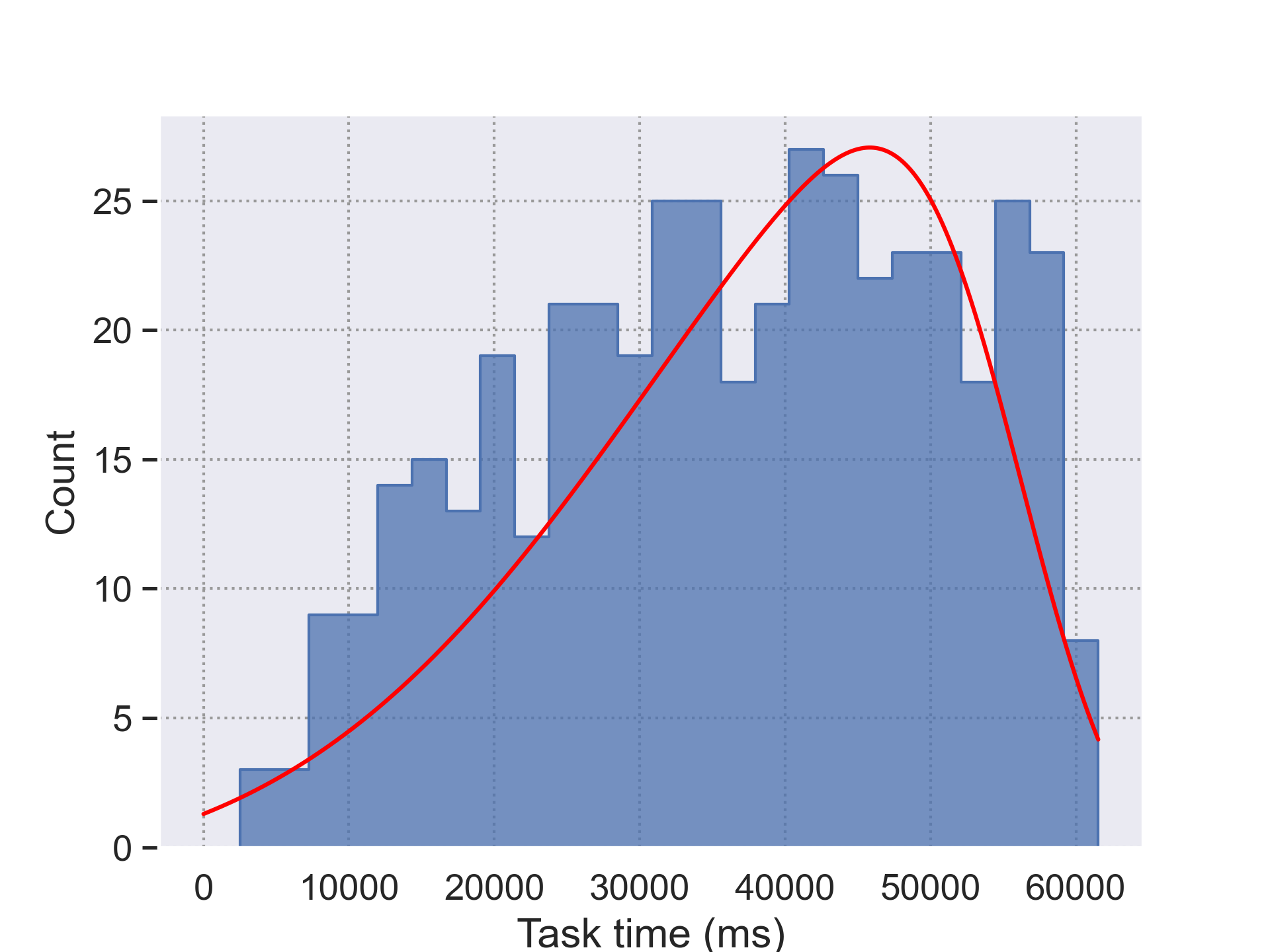}}
  \caption{Fitting result of Probability Density Function (red) for PBCs generated through the progress of 1-min verbal fluency task. It shows the frequency of PBCs given increases as the task proceeds, and decreases when the task is coming to an end.}
  \Description{This figure shows fitting result of Probability Density Function for PBCs generated through the progress of 1-min verbal fluency task as a skew normal distribution. It shows the frequency of PBCs given increased as the task proceeds, and decreases when the task was coming to an end}
  \label{fig:progressFitting}
\end{figure}

Where $Score_{pg}$ stands for Progress Score, k is the scaling factor, and $t_{task}$ is the time that a task has been proceeded. The PDF fitting result is shown in Figure ~\ref{fig:progressFitting}. Similar to our observation in the MoCA dataset, Progress Score peaked in the second half of task progress.

\subsubsection{Participant Score}
\label{section:Participant Score}

Participant Score was a score generated at the beginning of a test, describing the proactivity level of backchanneling for each participant i.e. how difficult to trigger a PBC.
Participant Score was motivated by the finding in \ref{sec:dataParticipants} that there was a clear individual difference among participants.
To simplify this question, we considered it as a probabilistic classification task, where the input was a segment of speech from the participant with fixed content and the output was a score. 
% \textcolor{red}{If a participant has a high Participant Score, the summed PBC score is higher, so it's easier to reach the threshold.}
% This scoring algorithm is based on the idea of Platt Scaling \cite{platt1999probabilistic}: the outputs of a classification model can be transformed into a probability distribution, by which the resulting probabilistic output can be used as a score. 
We first used SVM to classify participants into two classes: participants received more PBCs and participants received less PBCs. 
Then the output $d$ of SVM, i.e. distance from input data to SVM classification hyperplane, was used as a classification score.
Next, Platt Scaling \cite{platt1999probabilistic} was applied on $d$ to obtain a probabilistic value with range of (0, 1) from the output of SVM, denoted as Participant Score, through the following function:

% [comment previous function]
%\begin{equation}
%\label{eq:SubjScore}
%Score_{subj} = Platt(Dist_{svm}(x))
%\end{equation}

%where $Dist_{svm}(x)$ represents the distance from the input sample to the SVM %classification hyperplane, and

\begin{equation}
\label{eq:PlattScore}
Score_{subj} = Platt(y=1|d) = \frac{1}{1+exp({\alpha}f(d)+\beta)}
\end{equation}

where $Platt(y=1|d_{subj})$ refers to Platt Scoring, $d$ refers the output of classification model, $\alpha$ and $\beta$ are parameters learnt from SVM training data \cite{platt1999probabilistic}. Table~\ref{tab:SubjScore} shows the results of theoretical experiment using SVM to classify two groups of participants.
%It performs a linear transformation to the distance from input sample to SVM classification hyperplane, in order to obtain a probabilistic value with range of (0, 1) from the output of SVM.

%Firstly, the participants in the MoCA dataset were divided into two groups by the median of number of backchannels received, resulting two groups of participants received more PBCs and participants received less PBCs.

%In order to make the two classes more comparable, we used the speaker's speech in "Repeating Task", i.e. participants were asked to repeat a given sentence, to guarantee the content for the speech were the same. Then, following the methodology for RBC prediction algorithm, we applied the ComParE feature set and the same feature selection method to select proper input features, and fed those features into a SVM classification model. The function to derive Participant Score can be written as:

\begin{table}[]
\begin{tabular}{lllll}
\hline
\textbf{SVM} & \textbf{Acc} & \textbf{Pre} & \textbf{Recall} & \textbf{F1} \\
\hline
Valid & 0.623 & 0.611 & 0.624 & 0.605  \\
Test & 0.571 & 0.600 & 0.576 & 0.589 \\
\hline    
\end{tabular}
\caption{Results of SVM classification on two classes of participant.}
\label{tab:SubjScore}
% \centering
\end{table}

\subsubsection{Overall PBC Score}
\label{section:Overall PBC Score}
To summarize, Pause Score measured utterance-level timing, telling when PBCs should occur after participant's utterance, while Progress Score measured task-level timing, indicating when PBCs should occur within a task. Participant Score adjusted the proactive level to adapt to different participants. Then we calculated an overall PBC score through combining those three sub-scores with a weighted sum:

\begin{equation}
\label{eq:PBCScore}
Score_{PBC} = w_{pau}*Score_{pau} + w_{pg}*Score_{pg} + w_{pt}*Score_{pt}
\end{equation}

For the convenience of tuning parameter, we limited the sum of $w_{pau}$, $w_{pg}$ and $w_{pt}$ to be $1$, of which the range of $Score_{PBC}$ was (0, 1).
Then we set a threshold $thr_{PBC}$ to make PBC decisions if the $Score_{PBC}$ was beyond that threshold. The lower $thr_{PBC}$ was, the more likely it was for users to receive PBCs. In order to find optimized setups for user experience, the hyper-parameters $thr_{PBC}$, $w_{pau}$, $w_{pg}$ and $w_{pt}$ were tuned according to statistics and user feedback from piloting our algorithm with ten participants. For instance, $w_{1}$ could be turned down to increase the response time of PBC; $thr_{PBC}$ could be turned up to reduce the number of PBCs received by users.

\subsection{Implementation of Backchanneling Algorithms}

We integrated the two models into a fully functional system, which analyzed a user's speech data in real-time and provide RBCs and PBCs whenever appropriate.

% \subsubsection{Speaking Interval Detection}
Previous studies mainly considered a continuous prediction, i.e. continuously feeding input audio into a prediction model and deciding whether to give a backchannel at each frame \cite{ruede2019yeah, jain2021exploring, ortega2020oh}.
Voicing Probability was used as input feature in those methods, with an expectation that the prediction model was able to learn the correlations of backchannels and speaking intervals/utterances.
However, the uncertainties introduced by prediction models brought a risk of interrupting speakers in case of false alarms .

% \xm{may put this paragraph in related work instead. Here, just say that, to reduce the risk of ... we design...}.

In our system, instead of using Voicing Probability as a feature, we independently designed a Speaking Interval Detection (SID) module to recognize speaking utterance and speaking interval from the input audio.
First, Voice Activity Detection (VAD) was adopted to pre-process the input audio with a sliding window, aiming to label input frame as voiced or unvoiced frame. 
Next, in order to aggregate voiced/unvoiced frames and avoid noising prediction spikes, a delay trigger mechanism was used to further process VAD results, i.e. the audio segments were considered as \textit{speaking utterance} only when the number of speech frames detected was beyond a threshold; similarly, the segments would be considered as \textit{speaking interval} when the number of silence frames was beyond a threshold. 

With this method, speaking boundaries were visible to our system and RBC/PBC modules would only be triggered by speaking intervals.
% \sout{trying to avoid interrupting the speaker when he or she is talking. Such speaking intervals are not limited to pauses between long sentence breaks: VAD can catch relatively small gaps between words with a proper detection threshold. }
Besides reducing the risk of interrupting the speaker aggressively, this pre-processing method could lower the computation cost as well. It only passed speech data to backchannel prediction models when speaking intervals were detected rather than triggers those models recurrently. It could reduce the computation latency of our backchanneling pipeline, which was an important consideration of making backchannel decisions in real time.

\section{Evaluation of Proof-of-Concept System}
\label{sec:systemEvaluation}

To answer ``\RQT”, we developed a proof-of-concept system \textit{TalkTive} for speech-based NCD screening, with a multimodal interface and with the RBC and PBC modules embedded. Then we conducted a between-subject study to evaluate the performance and user experience of our \textit{TalkTive} system. The full system (Condition 2) was used to compare two conditions: Condition 0 (baseline condition) with the preset task functions and no backchanneling, and Condition 1 with the same task functions but providing RBCs. We obtained institutional IRB approval for the whole project prior to the study.

\subsection{System Architecture and Interaction Flow}

\begin{figure*}
\centerline{\includegraphics[width=0.8\linewidth]{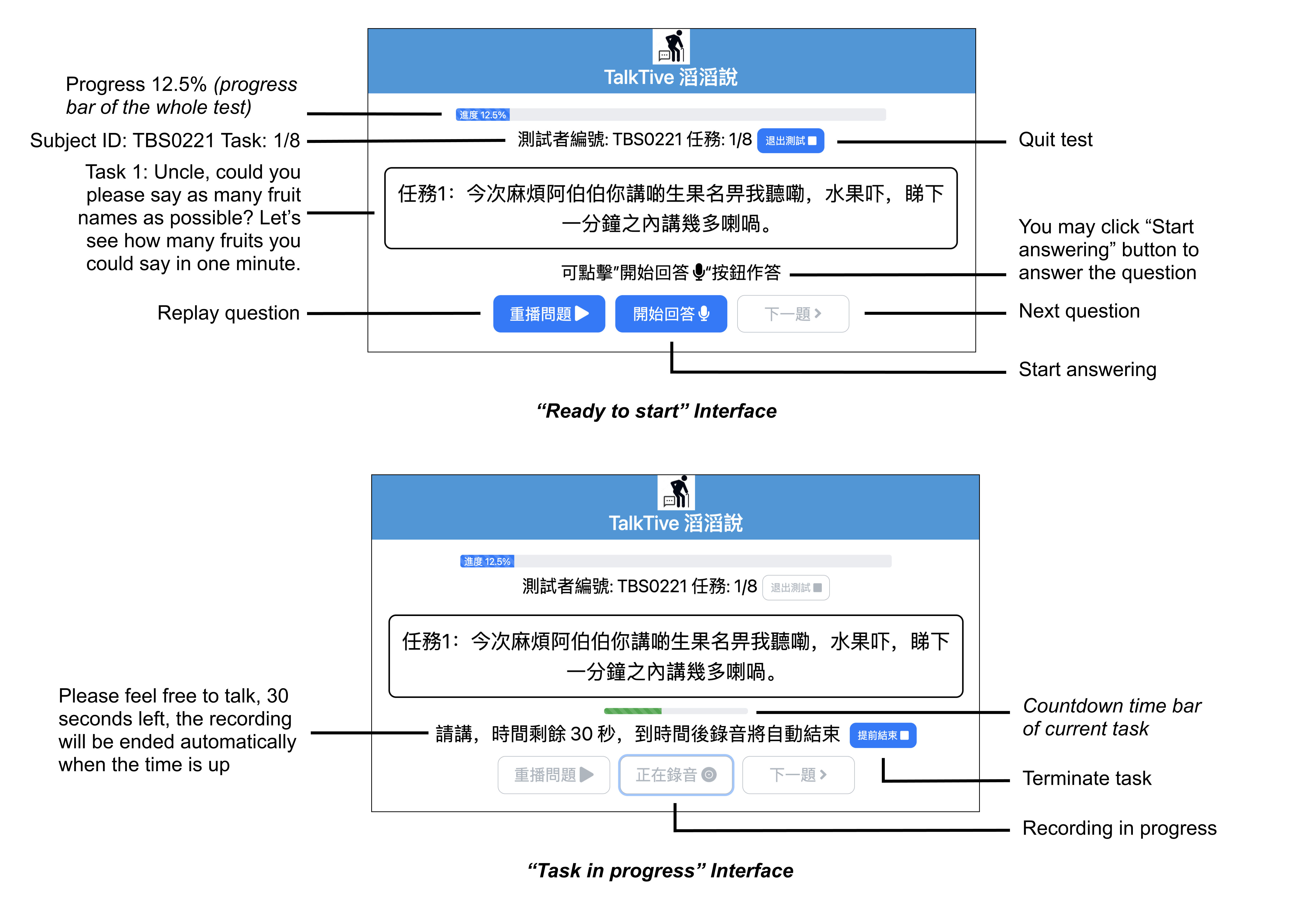}}
  \caption{\textit{TalkTive} interfaces: \textit{``Ready to start” Interface} (above), where the participant received task instructions and got prepared for answering; and \textit{``Task in progress” Interface} (below), where the participant provided answers in speech and could receive system-generated backchannels (Condition 1 \& 2).}
  \Description{This figure shows two interfaces of TalkTive: ``Ready to start” Interface, where the participant received task instructions and got prepared for answering; and ``Task in progress” Interface, where the participant provided answers in speech and could receive system-generated backchannels. The ``Ready to start” interface shows ``Progress 12.5\% (progress bar of the whole test)”, ``Subject ID: TBS0221 Task: 1/8”, ``Quit test” button, ``Task 1: Uncle, could you please name fruit names as many as possible? Let’s see how many fruits you could say in one minute.”, ``You may click “Start answering” button to answer the question”, ``Replay question” button, ``Start answering” button, ``Next question” button. The new things on “Task in progress” interface are ``Please feel free to talk, 30 seconds left, the recording will be ended automatically when the time is up”, ``Countdown time bar of current task”, ``Terminate task” button, and ``Recording in progress”.}
  \label{fig:interface}
  \vspace{-1mm}
\end{figure*}

\begin{figure*}
\centerline{\includegraphics[width=0.95\linewidth]{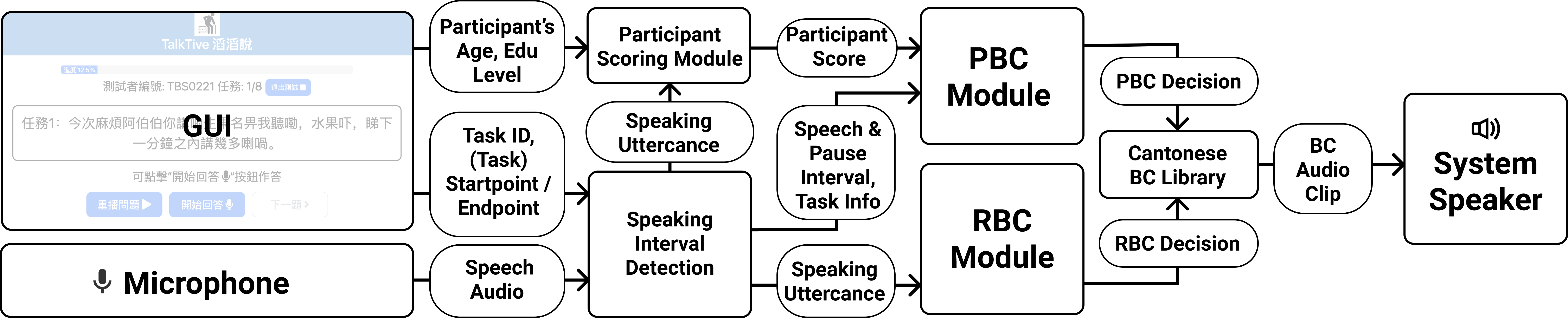}}
  \caption{\todo{The backend of \textit{TalkTive} system to predict} RBCs and PBCs during conducting cognitive assessments.}
  \Description{This figure shows how the backend of TalkTive process speech and give backchannel prediction. The frontend would send the participant's age, education level, task ID, startpoint & endpoint to the backend. Then the backend would process the audio as described in Figure 5, and select PBCs and RBCs in the Cantonese backchannel library.}
  \label{fig:workflow}
  \vspace{-4mm}
\end{figure*}

The \textit{TalkTive} system consisted of two parts: a React frontend as graphical user interface (GUI, see Figure~\ref{fig:interface}) and a Flask python server \todo{(see Figure~\ref{fig:workflow})}. Users interacted with the GUI to complete a series of MoCA tests as instructed by the conversational agent. In this process, the GUI issued corresponding commands to the backend as API calls, such as to open or close the RBC and PBC modules, and sent information inputted into the interface e.g., user's age to the backend. It also played back the speech output generated by the server.

More specifically, when a user came to the page of a new task, \textit{TalkTive} would introduce the task by playing an audio clip of task description. The user could replay the recording or start the task. For tasks with a time limit, a countdown would appear after the user clicks the ``Start answering” button. While the user was providing speech responses to the given question, \textit{TalkTive} ran the algorithms in Section \ref{sec:methodology} to provide RBCs and PBCs in real time.

To add the action module to the \textit{TalkTive} system, we invited a native Cantonese speaker experienced in conducting MoCA tests to record a Cantonese backchannel library. To build this library, we selected the most common RBCs and PBCs coded in the MoCA dataset, segmented related audio clips (which may also contain participant's speech and background noise), and passed them to an experienced MoCA assessor as prompts for audio recording. The assessor mimicked those audio clips of backchannels in a sound proof room. We indexed this Cantonese backchannel library and integrated into our backchanneling pipeline. If a RBC or a PBC was triggered, \textit{TalkTive} system would randomly play a piece of corresponding backchannel audio from the Cantonese backchannel library as feedback.

\subsection{Tasks}

There were three kinds of MoCA screening tasks, categorized based on the type of expected answer \ref{sec:dataTask} -- Type I questions asking for a one-off answer (e.g. ``please repeat the sentence…”), Type II questions prompting for a series of responses (e.g. ``please say animal names as many as possible in one minute”), and Type III open-ended questions. To ensure representativeness of experience in the user study, we sampled the three types of tasks (at least one question from each type) according to their frequency in MoCA. This resulted in a trial MoCA test of nine tasks: three Type I questions, five Type II questions, and one Type III question as listed below:

\begin{itemize}
\item (T0) \textit{Type I - Sentence repetition}: Please repeat the sentence that I said: [tongue twister in Cantonese]. (Trial task)
\item (T1) \textit{Type II - 1-min verbal fluency}: Please say fruit names as many as possible in one minute.
\item (T2) \textit{Type I - Sentence repetition}: Please repeat the sentence that I said: [tongue twister in Cantonese].
\item (T3) \textit{Type II - 1-min verbal fluency}: Please say animal names as many as possible in one minute.
\item (T4) \textit{Type II - 1-min verbal fluency}: Please say vegetable names as many as possible in one minute.
\item (T5) \textit{Type II - Serial 7 subtraction}: Please begin with 100 and count backwards by 7.
\item (T6) \textit{Type II - 1-min verbal fluency}: Please say place names in Hong Kong as many as possible in one minute.
\item (T7) \textit{Type III - Open-ended self-disclosure}: Could you please share a place you like and why?
\item (T8) \textit{Type I - Understanding}: How to say the word (clap) and how to perform that action?
\end{itemize}

In particular, T0 was designed to collect participants' speech and generate Participant Score, which was be used in the subsequent tasks. The method used to generate Participant Score was described in \ref{section:Participant Score}. T0 also served as a practice task to familiarize the participant with the \textit{TalkTive} system.

%We also found that backchannels happen more frequently in tasks asking for a series of responses and open-ended self-disclosure compared with those asking for an one-off response. And tasks requiring a series of responses are more prevalent than tasks requiring open-ended self-disclosure in the MoCA test. Hence we were most interested in how to provide backchannels in tasks requiring a series of responses. This resulted in a trial MoCA test of eight tasks: five tasks require a series of answers, one task requires an open-ended self-disclosure and two tasks require a one-off response. The task statements are attached in \todo{Appendix X}.

\subsection{Participants}

To conduct a between-subject user study with our target users -- older adults, we recruited \textit{n = 36} participants (19 females and 17 males, aged from 61 to 84 with an average of 72.4), 12 for each condition. We asked all participants to complete a pre-study survey based on their experiences in using electronic devices. According to their replies, all participants had a smartphone, 13 (36.1\%) had a tablet, and 13 (36.1\%) had a laptop or a desktop or both (one may own multiple devices). Most participants reported using electronic devices $3-10$ hours (17) and $1-3$ hours (12) per day. Only three participants (8.3\%) reported using electronic devices for less than one hour per day. These data suggested that our participants had daily access to electronic devices that could be used for CA-based NCD screening.

% Because our system was developed for NCD tests, we took subject's cognitive ability into our sampling process. There were three main groups for people who do not manifest serious NCD symptom and are the target subjects of NCD tests: healthy control (HC), health control with risk (HCR) and mild cognitive impairment (MCI). The cognitive ability level of participants for both conditions are comparable (baseline: 5 HCs, 3 HCRs and 4 MCIs; \textit{TalkTive}: 6 HCs, 2 HCRs and 4 MCIs). Hence we could control the effect of cognitive ability level of participants on the user study.

\subsection{Study Procedure}

The language used throughout the whole user study including the trial MoCA test was Cantonese. A researcher who was a native Cantonese speaker served as the experimenter and moderated the entire study process. At the beginning of the user study, the experimenter informed the participants that they would conduct a test of cognitive ability and memory. All participants signed a consent form, agreeing to join the research study and be audio recorded. After finishing a pre-study survey on participants' experience using electronic devices, participants were asked to watch a tutorial video of how to use the GUI as illustrated in Figure~\ref{fig:interface} by showing the steps to finish an example task (not used in the main study). After that participants were required to finish eight tasks independently, and the experimenter would not intervene unless there was a system breakdown. Upon completion of all eight tasks, the participants proceeded to fill out a post-study questionnaire on the same screen with assessor's facilitation (details in \ref{rec:postStudy}). They were required to rate their level of agreement to eight statements regarding their experience of talking to the (\textit{TalkTive} system or baseline system without backchannel), according to a 7-point Likert scale. We concluded the study with an exit interview of four questions about their user experience. We present the detailed survey and interview questions in the next subsection. %The entire study procedure was approved by our institution's IRB.

\subsection{Measures}

To evaluate the effect of backchannels provided by \textit{TalkTive} and user experiences of using the system for speech-based NCD screening, we collected a series of measurements from both experienced human assessor's and older adult participant's perspectives.

\subsubsection{Assessor's Validation of System-Generated Backchannel Responses}
\label{rec:assessorValidation}

We audio-recorded all the conversations between the participants and the \textit{TalkTive} system and marked all occurrences of backchannels generated based on system logs. We invited a trained assessor who has earned a certificate of running the MoCA test and also a native Cantonese speaker to evaluate the appropriateness of all backchannel instances given in Conditions 1 and 2. %The analysis from the trained assessor perspective was that a trained assessor who has earned a certificate of running the MoCA test and has native proficiency in Cantonese evaluates all given backchannels. 
%We logged all the occurrences of backchannels generated and recorded audio of the whole user study. 
The assessor listened to all task recordings and was prompted to rate each instance of backchanneling with binary choice -- inappropriate versus appropriate. This evaluation may be regarded as testing the precision of our backchanneling algorithms. Because of the optional nature of backchanneling \cite{ward2000prosodic, huang2010parasocial}, the selection and placement of backchannels may vary according to individualized preferences. Hence it was impractical to obtain the ground truth of all possible backchannels and identify the false negatives, so we did not include recall in the analysis.

\subsubsection{Post-study Questionnaire of Participants}
\label{rec:postStudy}

To obtain the participants' subjective feedback based their experience in interacting with our system, we adopted eight statements from \cite{cutrone2014cross}. The participant needed to provide a rating those statements (five from a positive perspective and three from a negative perspective) using a Likert scale from 1 (strongly disagree) to 7 (strongly agree). The five positive statements were ``the person who just asked me questions showed me that she understood what I said”, ``... she listened attentively to what I said”, ``... she encouraged me to talk”, ``... she was polite” and ``the test went smoothly”. And the three statements from a negative perspective were ``... she seemed impatient”, ``... she seemed cold and unfriendly” and ``... she interrupted me”. All participants were asked to provide a rating those eight statements.
% Because participants in Condition 0 wouldn't receive any response from the system, only participants in Condition 1 and 2 would be asked to rate those three statements as a comparison of user's experience on responses provided by the system.

We adopted the eight statements to gain a comprehensive understanding of the participants' experience with the \textit{TalkTive} system. We randomized the order of presenting the positive and negative statements to prevent the participants from simply giving the same rating to all items. To facilitate older adult participants to finish the post-study questionnaire thoughtfully, the experiment would read out each statement if participants had difficulty viewing the text by themselves. We also ensured that the participants fully understood the meaning of the scale. In addition, we invited the participant to give the reasons behind their ratings through thinking out loud, and we audio-recorded their verbal explanations.  %For the first three positive statements, each statement was followed by a negative statement to ensure that the participant would not simply give the same rating to all the statements.

\subsubsection{Semi-structured User Exit Interview}

After each participant finished the test and the post-study questionnaire, the experimenter conducted a semi-structured exit interview and audio-recorded the session with the participant's consent. The user interview has three main questions as listed below.

\begin{itemize}
\item Q1: How was your experience using this system? \textit{Follow up:} How did you feel about communicating with the person (the voice) who just asked you questions? 
\item Q2 [Asked in Condition 1 \& 2]: Did you notice the response given by the system, like ``hmm”, ``yeah”? How did you feel about them? Why?
\textit{Follow up:} Do you think those responses were different from those given by humans? If so, what was the difference?
\item Q3: Could you please give some suggestions to improve this system?
\end{itemize}

\section{Results from the User Study}
In this section we report on the quantitative and qualitative findings of our user study, covering the performance of the proposed backchannel-generation algorithms for NCD assessment, and the user perception of and experience with backchannel-enabled CA assessors to those without this mechanism.

% \xm{Any quantitative distribution data to show that the number distribution of our BCs was indeed adaptive to task and user characteristics? Was the distribution consistent with that in MoCA data? -- this was one unique characteristics of our system and it needed to be evaluated...}

\subsection{About 89\% of System-Generated Backchannels were Validated as Appropriate by Trained Assessor}

\begin{figure*}
\centerline{\includegraphics[width=1\linewidth]{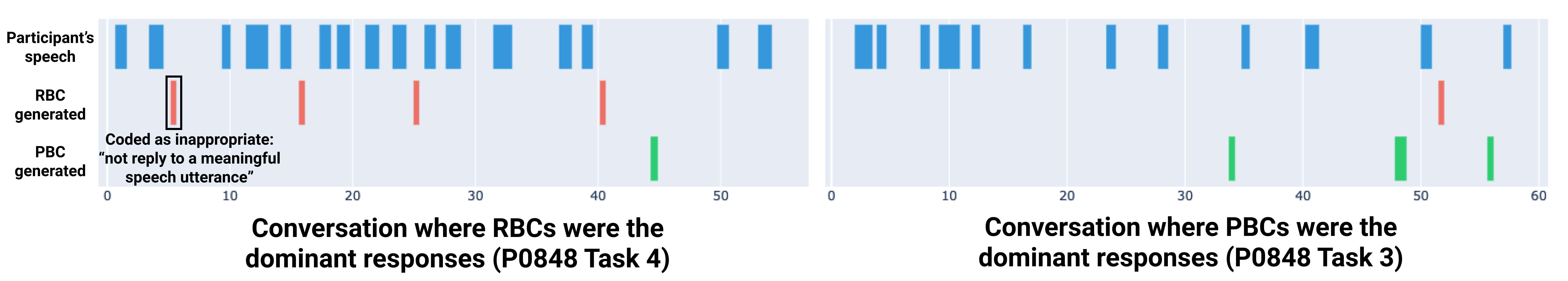}}
  \caption{Examples of conversation where RBCs were the dominant responses (P0848 Task 4) and conversation where PBCs were the dominant responses (P0848 Task 3). We could see that most RBCs directly follow speakers' utterances while PBCs mainly occur in longer pause. The first RBC in P0848 Task 4 was coded as inappropriate because it did not reply to a meaningful speech utterance. All the other system-generated backchannels were coded as appropriate.}
  \Description{This figure shows two examples of conversation where RBCs were the dominant responses (P0848 Task 4) and conversation where PBCs were the dominant responses (P0848 Task 3). In the example of RBCs, most RBCs directly follow speakers' utterances. In the example of PBCs, PBCs mainly occur in longer pause. The first RBC in P0848 Task 4 was coded as inappropriate because it did not reply to a meaningful speech utterance. All the other four system-generated RBCs and four system-generated PBCs were coded as appropriate.}
  \label{fig:bc_result}
\end{figure*}

Examples of the timing of RBCs and PBCs in conversations were shown in Figure \ref{fig:bc_result}. To verify that the timing and form of the BCs generated during the study conform to the common practice of human assessors, we invited an expert to classify each BC instance based on perceived appropriateness.
After we collected all the data of participants using the \textit{TalkTive} system, a trained assessor coded all 649 BCs (518 RBCs, 131 PBCs) given to the 24 participants in Condition 1 and 2 (details see \ref{rec:assessorValidation}). 

Participants in Condition 1 received a total of 267 RBCs (22.3 RBCs per test). Although more RBCs were generated by our system compared with the MoCA dataset (11.1 RBCs per test), most RBCs generated were considered as appropriate by expert: only 26 (9.7\%) of them were coded as inappropriate. A total of 251 RBCs and 131 PBCs were produced in Condition 2 (20.9 RBCs and 10.9 PBCs per test). The number of RBCs given was slightly lower than that of Condition 1, and the number of PBCs was comparable with the number of PBCs given in the MoCA test (8.28 PBCs per test). Among all BCs in this condition, 46 instances (12.0\%) were coded as inappropriate (see Table \ref{tab:assessor_code} for details), which demonstrated that our evaluator deemed the mechanism to be generally acceptable. %the validation result shows a general acceptance of backchannels from the perspective of assessor 
It seemed that the provided PBCs had a slightly higher chance than RBCs to be regarded as inappropriate, which aligned with the proactive nature of PBCs that they were encouraging but had the risk of being aggressive.

\begin{table*}[]
\begin{tabular}{|l|l|l|l|l|l|}
\hline
\textbf{Con} & \textbf{Type} & \textbf{\# of BCs} & \textbf{\# of Inappropriate BCs (\%)} & \textbf{Causes of Inappropriate BCs} \\ \hline
1 & RBC & 267 & 26 (9.7\%) & ``not reply to a meaningful speech utterance” (13/26); \\
& & & & ``interrupted speaker's thinking” (10/26) \\
& & & & ``interrupted speaker's talking” (1/26) \\
& & & & ``quicker than the participant's response” (1/26) \\
& & & & ``overlapped with `time's up'” (1/26) \\ \hline
& RBC & 251 & 27 (10.8\%) & ``not reply to a meaningful speech utterance” (24/27); \\
2 & & & & ``interrupted speaker's thinking” (3/27) \\ \cline{2-5}
& PBC & 131 & 19 (14.5\%) & ``urged the speaker” (10/19); \\
& & & &  ``interrupted speaker's thinking” (8/19); \\
& & & & ``not reply to a meaningful speech utterance” (1/19)\\
\hline
\end{tabular}
\caption{Results of assessor's validation of backchannels given in Condition 1 \& 2.}
\label{tab:assessor_code}
\end{table*}

To further investigate the causes of each inappropriate backchannel, we asked the assessor to take note of the reason behind her judgement. Overall, she specified three kinds of major causes (see Table~\ref{tab:assessor_code}). 1) ``Not reply with a meaningful speech utterance”. This mainly occurred when RBCs responded to the speakers' non-lexical utterances such as ``uhmm...” which did not include meaningful answers to acknowledge or agree with. 2) ``Urged the speaker”, which happened mostly when multiple PBCs were triggered within a short period of time. Also, 3) ``Interrupted speaker's thinking”, which frequently occurred for both RBCs and PBCs in the serial 7 subtraction task.

When reviewing the occurrences of inappropriate backchannels in the audio recordings, we found many cases happened while older adults were mumbling to themselves. For ``not reply to a meaningful speech utterance”, the muttering of older adults might be detected by the Speaking Interval Detection (SID) module and considered as a speaker utterance and then trigger RBCs. Figure \ref{fig:bc_result} shows a example. A similar situation may be observed for incomplete utterances with a long pause inside, which appeared frequently in the serial 7 subtraction task, as it was found to be difficult for the older adult participants.  For example, the participant might get the first digit much earlier than the second digit, such as ``eighty- [a long pause] six”. Human assessor would know that ``eighty” was a unfinished answer and would not backchannel to acknowledge that, while the pause might trigger our algorithm based on acoustic features to give a RBC. These kinds of situations indicated the need to use Automatic Speech Recognition (ASR) and Natural language processing (NLP) to understand the meaning of speakers' utterances, to identify the mumbling and also incomplete expressions, in order to yield more reasonable backchannels. However, currently there is a lack of high-performance ASR systems for older adult speech in Cantonese.
In other cases, older adults may murmur as they were thinking.  The SID may fail to capture such speech, and they system may treat it as a long silence and evoke PBCs, which may be considered considered intrusive for the speaker.

\todo{\subsection{System with Proper Backchanneling Feature were not Perceived as Disturbing by Older Adults}}

\begin{table*}[]
\begin{tabular}{|l|l|l|l|l|l|}
\hline
\textbf{Statements} & \textbf{Con 0} & \textbf{Con 1} & \textbf{Con 2} & \textit{\textbf{p-value}} & \textbf{\textit{H-value}} \\ \hline
\multicolumn{1}{|l}{\textit{Positive Statements}} & \multicolumn{1}{c}{} & \multicolumn{1}{c}{} & \multicolumn{1}{c}{} & \multicolumn{1}{c}{} & \multicolumn{1}{l|}{} \\ \hline
She understood what I said. & 6.0 & 5.0 & 5.0 & 0.23 & 2.90 \\ \hline
She listened attentively to what I said. & 6.5 & 5.5 & 7.0 & 0.33 & 2.20 \\ \hline
She encouraged me to talk. & 6.5 & 6.0 & 7.0 & 0.53 & 1.26 \\ \hline
She was polite. & 7.0 & 7.0 & 7.0 & 0.42 & 1.74 \\ \hline  
The test went smoothly. & 7.0 & 5.5 & 7.0 & 0.27 & 2.62 \\ \hline
\multicolumn{1}{|l}{\textit{Negative Statements}} & \multicolumn{1}{c}{} & \multicolumn{1}{c}{} & \multicolumn{1}{c}{} & \multicolumn{1}{c}{} & \multicolumn{1}{l|}{} \\ \hline
She seemed impatient. & 1.0 & 2.5 & 2.5 & 0.07 & 5.43 \\ \hline
She seemed cold and unfriendly. & 1.0 & 1.0 & 1.0 & 0.11 & 4.45 \\ \hline
She interrupted me. & 1.0 & 1.0 & 1.0 & 0.10 & 4.55 \\ \hline
\end{tabular}
\caption{Medians and chi-square test results of users' level of agreement to these statements on a Likert scale of 7 (Strongly Agree) to 1 (Strongly Disagree) for Condition 0, 1 and 2. No significant difference was observed for all the statements among three conditions.} %* are significant differences at \textit{p=0.05} level and ** are significant differences at \textit{p=0.01} level.
\label{tab:post-survey-3}
\end{table*}

To assess participants' subjective feedback on using our system, we compared the user ratings of eight statements in the post-study questionnaire across the three conditions. Because the answers of the post-study questionnaire were in the form of Likert-scale input, the data should be treated as ordinal measurements \cite{boone2012analyzing}. Hence we used the median to describe their central tendencies and applied Kruskal-Wallis non-parametric Test to evaluate the difference.  %To compare the rating on statements among condition 0, 1 and 2, we used Kruskal-Wallis Test as a non-parametric test for two or more groups. 
% We used Mann-Whitney Test to compare agreements on negative statements between Condition 1 and 2.

The statistical results were reported in Table \ref{tab:post-survey-3}. In general, there was no significant difference regarding agreements on the eight statements among three conditions. Although the results did not show that the system with backchannels outperformed the baseline condition in terms of encouraging speaker or listening to speakers attentively, they suggested that our system did not induce negative perceptions such as being less polite/patient/unfriendly after incorporating additional reactive or proactive backchanneling responses. However, regarding the central tendencies, Condition 1 with only RBCs  has the lowest ratings for five statements, indicating that RBCs only may tend to perform worse than no BCs or both RBCs and PBCs. % To further investigate the trend of each condition, we calculated effect sizes for each statement.

% \subsection{Qualitative Feedback on How Older Adults Perceived Backchannels}

\vspace{0.3cm}

\subsection{Qualitative Feedback on How Older Adults Perceived Backchannels}

In this section of qualitative analysis, the goal was to understand the participants' thoughts and comments on the backchannels. Condition 0 has no backchanneling, and hence we report on the users' feedback for Conditions 1 and 2.   there were no backchannels generated in Condition 0, here we mainly report the users' feedback on Condition 1 and 2.  

\subsubsection{Older adults reported that receiving only RBCs was not as good as gaining responses from human}

More than half of participants in Condition 1 (7 out of 12) reported that the responses generated by the system were not as good as those generated by humans. Four participants said that they did not even notice the responses generated by the system (P0877, P0316, P0304, P0215). P0316 stated that the RBCs were not articulate enough:

\begin{quote}

``I didn't notice that it was giving response to me. (actually 31 RBCs were given to her) Just asking me to finish one question and then next... At the beginning when I heard `hmm’, I had no idea about what it was doing. Really no idea. While after I heard two of that, I got to know that it was replying to me...Real human will be much warmer. I thought it (the system) was just a computer, a computer which could ask questions. It's different from face-to-face communication of human.” (P0316)

\end{quote}

It was reasonable for older adults to be insensitive to RBCs due to feeling nervous about the test, focusing on giving answers, even experiencing cognitive or hearing impairments. In other cases, although having noticed the existence of RBCs, P0215 stated that RBCs were rigid instead of human-like, and expressed a preference of in person communication: 

\begin{quote}

``There was no feedback. I noticed responses like `hmm', but I knew it's just recording. I didn't think it's real. Human assessors would be better. For human assessors I could see their facial expressions to know whether they were paying attention.” (P0215)

\end{quote}

Other participants explained why they thought human assessors would outperform that system regarding responses given: ``the responses generated by the system would be better than an inexperienced volunteer, but worse than an experienced assessor” (P0248); ``human assessors would give more explanations...had more interactions” (P0235); ``the responses given by humans would be clearer than those generated by the system” (P0280); ``the system was less friendly compared with human” (P0874) or simply ``the system had a lack of something compared with human” (P0877).

For the remaining five participants in Condition 1, only P0306 said the responses generated by the system were better than those from humans because some people may not give any responses and just wait: ``The response was pretty good. It indicated that my answers were acceptable. The responses generated by the computer were better than those of human. Human would not give as much as response - he or she might just wait for you. Human might even not give response to you. Just let you think alone.” The other four participants said that the responses generated were similar to human's without giving further explanation.

\subsubsection{PBCs were appreciated by older adults, especially when they ran out of answers and were about to give up }

%While comparing with subtle negative qualitative feedback on Condition 1 which only had RBCs, 
Users gave more positive feedback on Condition 2 and they appreciated the existence of PBCs. In Condition 2, 8 out of 12 participants expressed that PBCs were well received by them. For example, P0250 found that PBCs were quite encouraging when she ran out of answers:

\begin{quote}
``(The sound of) AI was quite machine-like, while this (system) was not. (I) think it was operated by a human. For example, just now when I had not answered, it was listening to me carefully, and encouraging me in the meantime, just like doing Q\&A with you in person. It felt like that. We were nervous (in this kind of assessment), and it seemed that my mind suddenly went blank just now. (My brain) suddenly stopped (thinking). Could not think anymore. But it said `feel free to keep thinking'. It listened to me quite attentively.” (P0250)
\end{quote}

In Condition 2, it seems that all participants noticed the system-generated backchanneling, in contrast with Condition 1 where four participants stated that the system-generated responses went unnoticed. One participant, P0221, even compared system-generated response with the audio recordings of task statements, and she found the responses sounded more natural than task statements recorded by the human assessor. She stated the audio recordings of statements were machine-like, but the responses were not. She said, ``While I was thinking, (it said) `uhm', `keep going', so I would really think (about the answers), much better than if it didn't give any response...because it made me feel like that there was really a person there, `take your time', not like facing a cold computer”.

When the participants were asked about the difference between the responses provided by the system compared with those from a human, P0232 said that the response was a good signal of listener's attentiveness, which might not even  be provided by human. For example, P0232 and P0854 claimed that the system performed better than human and they felt more comfortable in talking to the system than with a human.:

% She thought the system was listening carefully, while she knew her memory was not good and she got used to people being impatient, so she gave a high rating (6) on ``she seemed is impatient” and low rating (3) on ``the test process is smooth”.

\begin{quote}
``(The responses from the system) showed that it valued my answers. Sometimes a person said `you speak', but there was still no response even after I finished. It seemed that he didn't like that you spoke to him, didn't want to listen to you. If it had response, (it meant that it) paid attention to your talk.” (P0232)
\end{quote}

\begin{quote}
``The responses were good. It would not make you feel nervous... Responses from a real human made me felt more nervous and stressed... (I was) more comfortable to talk to a computer (than a human)... The response from the system was encouraging you to keep talking, while the response from a system was more inflexible compared with human.” (P0854)
\end{quote}

\section{Discussion}

\subsection{Result Summary}
In general, there was no significant difference among conditions with system-generated backchannels and the baseline condition with no backchannel given in either a positive or a negative way. This may imply that the participants did not feel overly anxious after receiving system-generated backchannels, even though some instances might be deemed intrusive or aggressive from a professional assessor's point of view. Clearly providing appropriate backchannels is still far away from providing ``optimal” backchannels to help participants stay in the optimal arousal level, while eliminating the portion of inappropriate backchannels may be a concrete next step.

We found that reactive backchannels (RBCs), of which 89.8\% were coded as appropriate by expert, may be ignored by older adults or considered as rigid. On the other hand, although proactive backchannels (PBC) have a higher risk of being perceived as pushy or intrusive by the expert (14.5\%) than RBCs (10.2\%), they were well received by most older adult participants in the given task setting. We observed in our study that older adults, especially those who were experiencing a decline in cognitive ability, expected the system to provide articulate and noticeable instructions and responses to guide them through the tasks. This preference of older adults was reflected as a general acceptance of PBCs. As shown in the user interview, older adults tended to consider receiving PBCs as being encouraged instead of being urged. Particularly, PBCs could be more helpful when participants intended to give up early, which was common in a cognitive assessment setting.

\subsection{Design Considerations}
Based on the above findings, we propose a set of design considerations for future improvement of conversational agents (CAs) conducting cognitive assessments with backchanneling function. 
\todo{First, we need to ensure that the generated speech of the task-oriented CA fits the content and nature of the tasks specifics, i.e., administrating a neurocognitive disorder screening test. As postulated by the well-known Yerkes–Dodson law \cite{yerkes1908relation},  task performers should have an optimal mental arousal to perform a certain task. If the arousal of the task performers is lower or higher than the optimal level, their performance may be hindered by inactivity or anxiety. To guarantee that the participants' performance may fully reflect their actual cognitive abilities in the MoCA test, we hope to situate them in an appropriate arousal level and engage them in the task without making them feel too stressful. As shown in the qualitative results of the user study, PBCs were deemed effective by over half of participants of Condition 2 in keeping them engaged. In comparison, only a single participant responded as such for RBCs. Participants also expressed that PBCs were showing attentiveness instead of stressing them out. In the qualitative feedback. Hence, it may be beneficial to include PBCs to boost the task performers' arousal level so that they stay active in the process. %some participants found RBCs were unnoticeable and PBCs were observed to be more effective to engage participants. 
At the same time, as suggested by the clinical assessors, we need to control the frequency and intensity of PBCs to avoid pushing the users beyond their normal arousal stage, which may been undesirable for cognitive assessment.}
%so the participants did not feel over anxious which had a negative effect on both authenticity and experience of this cognitive assessment.

%As shown in the results the post-study questionnaire, currently there is no significant difference among conditions with system-generated backchannels or baseline condition with no bakchannel given in either a positive or negative way. It revealed that the participants were not feeling over anxious after receiving system-generated backchannels in general, while some certain backchannels might be considered as intrusive or aggressive which were coded by expert as ``urged the speaker” and ``interrupted speaker's thinking”. Clearly providing appropriate backchannels is still far away from providing ``optimal” backchannels to help participants stay in the optimal arousal level, while eliminating the portion of inappropriate backchannels could be the next step.

Second, for improving the precision of system-generated BCs, reducing the number of inappropriate PBCs may be prioritized because PBCs are more noticeable and might affect participants' arousal more significantly than RBCs. According to the Yerkes–Dodson law, difficult or intellectually demanding tasks may require a lower level of arousal to facilitate concentration for optimal performance \cite{yerkes1908relation}. This may explain why PBCs in the serial 7 subtraction task are often coded as inappropriate by the expert (10/19). To resolve the tension between encouraging versus pushing the participants while they perform intellectually demanding tasks, we may further investigate task-related adaptivity. Based on analysis of the MoCA dataset and results of user study, we may extend the cooling time of providing continuous PBCs in those difficult tasks to avoid ``urged the speaker”, as well as improve the  Speaking Interval Detection (SID) module to detect murmur by the older adult participant and hence avoid interrupting their thinking process. Improvement of SID may also improve to RBC generation in a similar manner. We observed only one instance of inappropriate PBCs  caused by the inability of the system to understand the speaker' utterance, which suggests that the use of Automatic Speech Recogntion (ASR) may not be essential for triggering PBCs.
%system failing to understand the meaning of speaker's utterance, further enhancing the reliability of Automatic Speech Recognition (ASR) algorithm is not essential to lower the occurrences of inappropriate PBCs.%, since it can prevent our algorithm from interrupting speaker's thinking and talking as well. 
%It shows the potential of our backchanneling strategies to keep reducing the ratio of inappropriate backchannels.

Third, aside from reducing instances of inappropriate backchannel, we also need to improve participant-related adaptivity of backchanneling strategies. According to the Yerkes–Dodson law, each task performer has an optimal mental arousal to perform a certain task \cite{teigen1994yerkes}. Clearly there exist individual differences among participants in terms of skill level (cognitive ability), personality and trait anxiety as potential influencers of arousal level \cite{matthews1996attentional, friedman1979framing, berlyne1960conflict}. As a result, the optimal arousal level may vary from person to person. %which causes the optimal arousal and external response required to help them achieve that optimal also vary from person to person. 
In addition, individual difference in sensitivity may explain why some participants did not notice RBCs while others did. We have attempted to integrate the idea of adaptive backchanneling based on participant characteristics encoded as \textit{Participant Score}. This was a first step towards finding an optimal, personalized backchanneling strategy for each user. To develop such adaptive strategies, we will need to further conduct within-subject studies with multiple levels of proactivity in backchanneling. Other information about users e.g., language \cite{heinz2003backchannel, clancy1996conversational}, culture \cite{cutrone2014cross}, etc. may also affect how users perceive backchannels, and thus may be incorporated into the personalized backchanneling strategy. %, and we will keep investigating the implementation and evaluation of adaptive backchanneling in the future.

\subsection{Generalization}
Although our work was evaluated in a specific task scenario for a particular user population, it may be generalized in three ways. First, the backchanneling algorithm we propose may be applied to other kinds of task-oriented conversations requiring speech responses, including other cognitive tests \cite{hoops2009validity}, self-disclosure \cite{leaper1995self}, counseling \cite{kawahara2016prediction}, to name a few. While the timing and frequency of backchanneling, especially for PBCs, may need to be adjusted according to the social rules and common human-human interaction practices in the target tasks, our data-driven approach to modeling backchannels has the potential to be easily adapted. Second, our CA system may be extended to serve other user groups, e.g., young adults who are vulnerable to neurocognitive disorders \cite{panegyres2007course}, children who need to be engaged in learning tasks \cite{park2017telling}, etc. Third, we demonstrated a workflow of investigating backchanneling patterns in Cantonese as a low-resource language. Different from well-studied languages like English, it is a challenge to study backchanneling for a language without previous work on this topic and supportive toolkits such as reliable ASR algorithms. Our work introduces potential solutions to overcome those constraints in both data analysis and implementation strategies, which might be beneficial for future research on backchanneling targeting low-resource languages.

\subsection{Limitations and Future Work}
There are several limitations regarding the design of the system and the experiment.
As a proof-of-concept system, the TalkTive system is currently deployed on a desktop. We plan to develop a mobile APP version of it in the future, which offers accessibility via mobile phones.  Also, the current study was conducted in the laboratory with a trial version of the MoCA test.  The full version of the MCA test can be conducted with the mobile APP in the future. Another limitation associated with the in-lab setup is the difficulty in recruiting older adult participants due to their reduced mobility especially under the restrictions under COVID-19. \todo{Compared with the size of potential user population of the \textit{TalkTive} system (i.e., the number of older adults in Hong Kong), the evaluation study was underpowered with 36 participants divided into three conditions.} With more participants, our study will have a higher statistical power in analyzing the differences among various backchanneling strategies \todo{and revealing the general preference of older adults}. With more participants, we will also be able to further analyze the relationship between the characteristics of the participants and their acceptance of backchanneling.  This will help achieve greater personalization of backchanneling strategies.  In addition, if a high-performance ASR system for older adult speech becomes available, the system may better understand the user's utterance and improve the trigger of RBCs.
%\todo{Apart from underpowered sample size, the current system deployment was in a lab setting and not in the wild, and only in one time point, under researcher supervision.} Besides, a lack of mature ASR algorithm for older adults' speech in Cantonese also limited the performance of our backchanneling algorithms, especially RBCs which were responding to speakers' meaningful utterances.

\todo{Moreover, there still exist intrinsic limitations of CAs as compared with human healthcare professionals, such as not being as responsive and empathetic \cite{kim2019conversational, kretzschmar2019can, silverman2016artificial}. We believe that human therapists and CAs are complementary to each other and CAs should not aim to replace human healthcare professionals. In this study we have demonstrated that CAs present an affordable solution that enables older adults to take pre-screening tests in their homes and communities. For users considered to have a high risk of neurocognitive disorder based on pre-screening results, they need to seek further screening, diagnoses and treatment in the clinic. In sum, CAs have the potential in offering a scalable, accessible and economical means to support preliminary assessment of cognitive decline, and can help support the clinician's work to attain greater effectiveness and higher efficiency.}

\section{Conclusion}

Conversational agents (CA) hold a strong promise in supporting digital cognitive assessments, with minimal human intervention, in order to scale up cognitive screening for early detection of NCDs. This paper presents an approach for automatic generation of backchanneling, a type of verbal response that enables the CA to acknowledge the user's input and encourages them to interact further. We analyzed a dataset with 246 human-human conversations involving an assessor and a participant in a the Montreal Cognitive Assessment (MoCA) test. We identified two kinds of backchannels – reactive backchannels (RBCs, e.g. ``hmm”) and proactive backchannels (PBCs, e.g. ``what's more”) – commonly adopted by human assessors. We labeled 2,732 RBCs and 2,037 PBCs in the MoCA dataset, and devised a data-driven method to model the timing of the two types of backchanneling. We proposed algorithms that generate RBCs and PBCs based on task-related and participant-related patterns, and developed TalkTive, a CA which can predict the timing and form of backchanneling while conducting speech-based cognitive assessments. For evaluation, we conducted a between-subject, in-laboratory study with n=36 older adult participants. The study demonstrated that the backchanneling algorithm can effectively generate PBCs and RBCs, over 88\% of which were deemed appropriate by a human expert. Quantitative and qualitative evaluations in participant experience reflected that the automatically generated backchanneling was regarded as smoothly incorporated and not intrusive, while PBCs were preferred to RBCs.
\begin{acks}
\todo{This project was supported by the HKSARG Research Grants Council's Theme-based Research Grant Scheme (Project No. T45-407/19N). We would like to thank Pauline Kwan for recording a Cantonese MoCA test and backchannel library, and Stephen MacNeil for feedback. We would also like to thank all the study participants for their time and feedback.}
\end{acks}

\bibliographystyle{ACM-Reference-Format}
\bibliography{sample-base}

% \appendix

% \section{Appendix A}

% \section{Appendix B}

\end{document}